\documentclass[10pt,twocolumn,letterpaper]{article}

\usepackage[cvprfinal]{cvpr}      % To produce the REVIEW version
\usepackage{times}
\usepackage{epsfig}
\usepackage{graphicx}
\usepackage{amsmath}
\usepackage{amssymb}
\usepackage{booktabs}
\usepackage{multirow}
\usepackage{color}
\usepackage{makecell}
\usepackage{xcolor}
\usepackage{caption}
\usepackage{bbding}
\usepackage{pifont}
\definecolor{cvprblue}{rgb}{0.21,0.49,0.74}
\usepackage[pagebackref,breaklinks,colorlinks,citecolor=cvprblue]{hyperref}

\usepackage[capitalize]{cleveref}
\crefname{section}{Sec.}{Secs.}
\Crefname{section}{Section}{Sections}
\Crefname{table}{Table}{Tables}
\crefname{table}{Tab.}{Tabs.}

\def\paperID{3705} % *** Enter the Paper ID here
\def\confName{CVPR}
\def\confYear{2024}

\title{Neural Super-Resolution for Real-time Rendering with Radiance Demodulation }
\author{
Jia Li\textsuperscript{1}, Ziling Chen\textsuperscript{1}, Xiaolong Wu\textsuperscript{1}, Lu Wang\textsuperscript{1,*}, Beibei Wang\textsuperscript{2,3,*}, Lei Zhang\textsuperscript{4}
\and 
\textsuperscript{1}Shandong University, 
\textsuperscript{2}State Key Laboratory for Novel Software Technology, Nanjing University,
\and
\textsuperscript{3}School of Intelligence Science and Technology, Nanjing University,
\and
\textsuperscript{4}The Hong Kong Polytechnic University
 \\ {\tt\small luwang\_hcivr@sdu.edu.cn}, {\tt\small beibei.wang@nju.edu.cn}
}

\begin{document}

\twocolumn[{
\renewcommand\twocolumn[1][]{#1}%
\maketitle
\begin{center}
\centering
\includegraphics[width=1.0\textwidth]{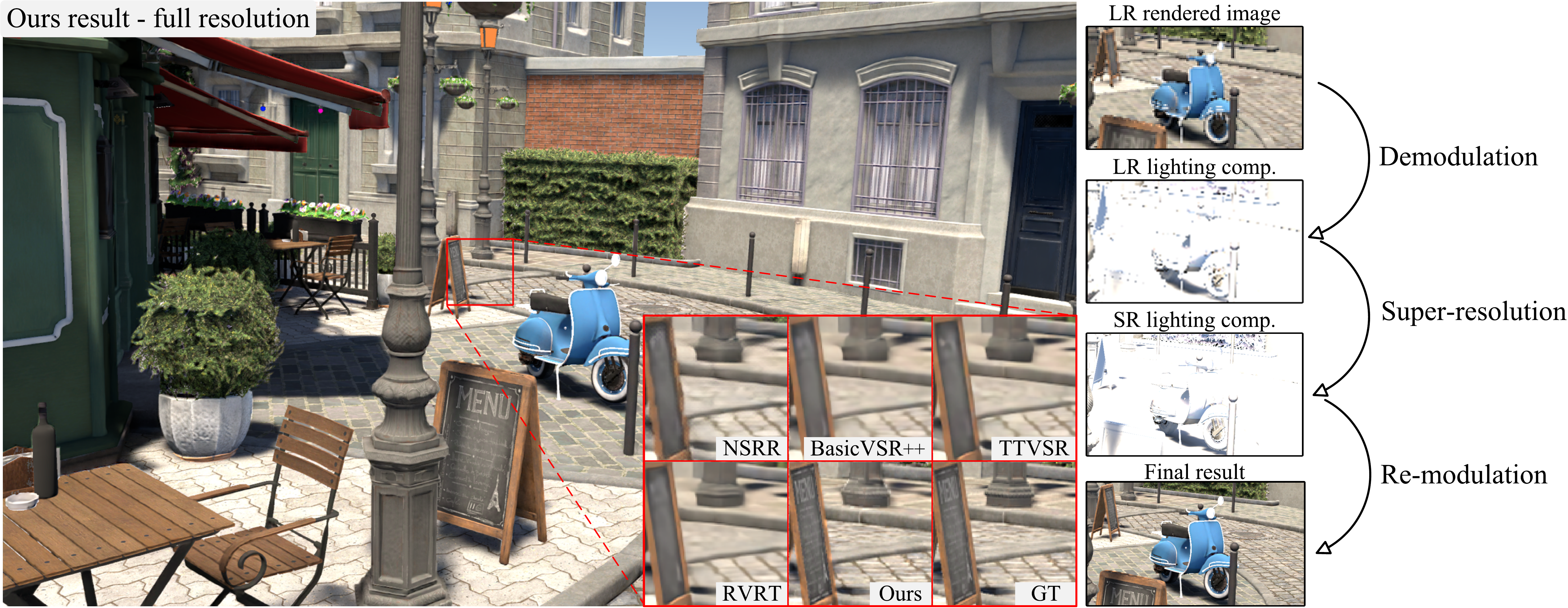}
\captionof{figure}{
Quality comparison between our method and the state-of-the-art methods NSRR~\cite{NSRR}, BasicVSR++~\cite{chan2022basicvsr++}, TTVSR~\cite{liu2022ttvsr} and RVRT~\cite{liang2022rvrt}. The upsampling ratio is set as 4 $\times$ 4. By radiance demodulation, we perform super-resolution on the smooth lighting component only, allowing our method to preserve richer scene details after re-modulation with the high-resolution material component.}
% Quality comparison between our proposed method and the other four methods on Bistro and Pica scenes, with the upsampling ratio set as 4 $\times$ 4. Our method significantly outperforms the existing state-of-the-art methods both visually and quantitatively, preserving more details and reducing ghosting artifacts (pointed out by the red arrow).}
\label{fig:teaser}
\end{center}
}]

\renewcommand{\thefootnote}{*}
\footnotetext{Corresponding author.}

\begin{abstract}
It is time-consuming to render high-resolution images in applications such as video games and virtual reality, and thus super-resolution technologies become increasingly popular for real-time rendering. 
However, it is challenging to preserve sharp texture details, keep the temporal stability and avoid the ghosting artifacts in real-time super-resolution rendering. To address this issue, we introduce radiance demodulation to separate the rendered image or radiance into a lighting component and a material component, considering the fact that the light component is smoother than the rendered image so that the high-resolution material component with detailed textures can be easily obtained. We perform the super-resolution on the lighting component only and re-modulate it with the high-resolution material component to obtain the final super-resolution image with more texture details. A reliable warping module is proposed by explicitly marking the occluded regions to avoid the ghosting artifacts. To further enhance the temporal stability, we design a frame-recurrent neural network and a temporal loss to aggregate the previous and current frames, which can better capture the spatial-temporal consistency among reconstructed frames. As a result, our method is able to produce temporally stable results in real-time rendering with high-quality details, even in the challenging 4 $\times$ 4 super-resolution scenarios. Code is available at: \href{https://github.com/Riga2/NSRD}{https://github.com/Riga2/NSRD}. %, and outperforms the existing state of the arts by an average of 0.82 dB in PSNR on our test scenes.

\end{abstract}

% \begin{figure}
%   \centering
%    \includegraphics[width=1.0\linewidth]{figures/new/teaser3.pdf}
%    \caption{4 $\times$ 4 super-resolution results on Pica scene. We proposed a novel real-time method by performing super-resolution on the texture-free weighted-lighting component only. It significantly outperforms existing state of the arts) by recovering more scene details and reducing ghosting artifacts.}
%    \label{fig:teaser}
% \end{figure}

%%%%%%%%% BODY TEXT
\newcommand{\tabincell}[2]{\begin{tabular}{@{}#1@{}}#2\end{tabular}}

% \begin{table*}
%   \centering
%   \small
%   \setlength\tabcolsep{2pt}
%   \caption{Comparison between VSR and RRSR. \textit{Auxiliaries} indicates whether auxiliary buffers are available, \textit{Alignment} indicates the spatial transformation applied to the misaligned images/features, and \textit{Propagation} means the way of temporal feature propagation.}

%  % \added{We need to explain something here, similar to the BasicVSR. We do not even explain, what's alignment and propagation.}}
%   \label{tab:VSR-RRSR}
%   \begin{tabular}{l|c|c|c|c|c}
%     \hline
%      Method & Type &  Real-time speed & Auxiliaries & Alignment & Propagation \\
%     \hline
%     BasicVSR++~\cite{chan2022basicvsr++} &VSR & \XSolidBrush  &\XSolidBrush &  Optical Flow  &  Bidirectional           \\
%      NSRR~\cite{NSRR}  & RRSR & \Checkmark & \Checkmark &  Motion Vector  &  Unidirectional  \\
%   \hline
% \end{tabular}
% \end{table*}
\vspace{-1.5em}
\section{Introduction}
\label{sec:intro}

%Rendering creates 2D images from virtual 3D scenes by simulating the light propagation in the scene. One important category of rendering is real-time rendering, which is widely used in real-time applications, like video games and virtual reality. Obviously, real-time rendering emphasizes the rendering time cost. One factor which impacts the time cost is the image resolution, regardless of the underlying rendering pipeline. Hence, real-time rendering usually tries to avoid rendering a high-resolution image directly. An alternative way is rendering a low-resolution (LR) image and then performing super-resolution (SR).

Real-time rendering is widely used in various applications, like video games and virtual reality, where both low time cost and high resolution images are desired. To achieve such a goal, super-resolution (SR) rendering technologies have been popularly adopted by first rendering a low-resolution (LR) image and then performing SR on it. However, SR for real-time rendering is challenging since it needs to meet several requirements: detail-preserving, temporally stable, artifacts-free, and highly efficient. 

Many methods have been developed for a relevant task -- video super-resolution (VSR)~\cite{chan2021basicvsr, chan2022basicvsr++, tecoGAN, isobe2020rrn, DUF, liang2022vrt, EDVR, liu2022ttvsr, liang2022rvrt}. These methods have shown impressive results but can not be used in real-time rendering super-resolution (RRSR). First, most of them rely on a heavy network for optical flow estimation~\cite{ranjan2017spynet}, resulting in an expensive time cost. Furthermore, these methods (e.g.,~\cite{chan2021basicvsr, chan2022basicvsr++, DUF, FRVSR, EDVR, liang2022vrt}) require both the precedent and the subsequent frames for bidirectional propagation. Unfortunately, only the precedent frames are available, since SR is performed simultaneously with rendering. 

Another line of works~\cite{FSR, TAAU, NSRR, guo2022classifier, mercier2023efficient} focus on RRSR. These methods exploit auxiliary buffers (e.g., depth buffer) to aid SR and can achieve real-time speed. However, due to the limited texture information in the LR image, even with these auxiliary buffers as inputs, the network can not recover the missing details, leading to blurry results. Furthermore, these methods have ghosting artifacts, when the motion vector~\cite{MV} becomes unreliable for occluded regions in dynamic scenes. These artifacts can be alleviated by predicting regions with networks~\cite{NSRR, guo2022classifier}, but still exist.  
%Some of them~\cite{NSRR, guo2022classifier} predict these  using networks without any explicit clues.}
%Instead, we propose a novel method to explicitly and accurately compute the unreliable regions.}

In this paper, we resolve these two issues with simple solutions. First, we introduce radiance demodulation into SR, together with a formulation to enable non-diffuse materials, inspired by real-time rendering denoising~\cite{pre-in}. Two key observations that make radiance demodulation feasible in SR are: 1) the rendered image (radiance) can be demodulated into a material component with rich texture details and a relatively smooth lighting component compared to the radiance; 2) the material component can be captured quickly, even capturing a high-resolution (HR) one directly. More specifically, we design a demodulation module that separates the lighting component from the rendered image and performs SR on the smooth lighting component only, while the HR material component is used directly for re-modulation with the SR results. In this way, we can obtain results with rich texture details. To our knowledge, we are the first to use radiance demodulation in RRSR. Second, we propose a simple way to get an occlusion-aware motion mask by subtracting two types of motion vectors, which can explicitly and accurately characterize the unreliable region for the network. Hence, the ghosting artifacts can be avoided. Finally, we design a lightweight frame-recurrent neural network using a convolutional long short-term memory (ConvLSTM) module, together with a temporal loss, which fully utilizes the intermediate features between adjacent frames to enhance the reconstruction quality and temporal stability further. Our method is specialized for real-time rendering and significantly outperforms prior work, including state-of-the-art RRSR and VSR methods, both visually and quantitatively.

%\revise{Based on these observations, we design a demodulation module that separates the lighting component from the rendered image and performs SR on the smooth lighting component only, while the HR material component is used directly for re-modulation with the SR results. In this way, we can obtain results with rich texture details. To the best of our knowledge, we are the first to use radiance demodulation in RRSR, even though a similar idea (only dividing the diffuse albedo, which only works for diffuse materials) has been widely used in denoising~\cite{KPCN, schied2017svgf, kontkanen2004irradiance}.} Regarding the ghosting artifacts, we design a novel occlusion-aware motion mask based on two kinds of motion vectors, explicitly and accurately pointing out the unreliable region for the network so that the ghosting artifacts can be avoided. Integrating the above techniques, \revise{we further enhance the reconstruction quality and temporal stability with a designed lightweight frame-recurrent neural network and temporal loss. Our method is specialized for real-time rendering and significantly outperforms prior work, including RRSR and state-of-the-art VSR methods, both visually and quantitatively.} 

To summarize, our main contributions include:
\begin{itemize}
    \item we introduce radiance demodulation into super-resolution for non-diffuse materials to better rich texture details,
    \item we propose a simple way to get an occlusion-aware motion mask, which avoids the ghosting artifacts in dynamic scenes, and
    \item we design a lightweight frame-recurrent neural network for real-time SR, together with a temporal loss, to improve the reconstruction quality and the temporal stability.
\end{itemize}

\section{Related Works}
\label{sec:related}

\paragraph{Video Super-Resolution.} 
%VSR reconstructs the current frame with multiple frames as input, and considers temporal consistency at the same time. 
Existing VSR methods can be roughly categorized into sliding window-based~\cite{VESPCN, tgt, li2020mucan, tian2020tdan, EDVR} and recurrent-based~\cite{huang2015rcnn, FRVSR, haris2019rcnn, fuoli2019RLSP, RSDN, isobe2020RTM, lin2021fdan, chan2021basicvsr, chan2022basicvsr++}. The sliding window-based methods often take a short sequence of low-resolution frames as input without considering the correlation between reconstructed frames. By introducing the previously reconstructed images, recurrent-based methods have better temporal coherence. Huang et al.~\cite{BRCN} are the first to introduce a recurrent neural network in VSR by connecting hidden layers between adjacent frames with recurrent convolutions. Later, Sajjadi et al.~\cite{FRVSR} propose to warp the previously reconstructed frame into the network, which is further improved by Chan et al.~\cite{chan2021basicvsr, chan2022basicvsr++} with enhanced propagation and alignment. Recently, transformer-based structures have been introduced into VSR problems (Liang et al.~\cite{liang2022vrt, liang2022rvrt}). And Liu et al.~\cite{liu2022ttvsr} propose a trajectory-aware transformer to learn spatio-temporal information more effectively. All of them can achieve good results, at the cost of large parameters and expensive time cost.

Frame alignment is an important operation in VSR. Most previous methods~\cite{VESPCN, liao2015video, kappeler2016video} utilize optical flow to warp the previous frames. Another group of methods uses 3D convolution~\cite{DUF} or non-local methods ~\cite{yi2019progressive} to extract spatial-temporal information, without performing explicit frame alignment. Thus, they all need to design a more complex network for effective feature extraction.

% VSR has shown impressive results but can not be exploited for real-time rendering, as shown in Table~\ref{tab:VSR-RRSR}. However, some networks indeed inspire the design of our network, like the recurrent framework.
VSR has shown impressive results but can not be exploited for real-time rendering. However, some networks indeed inspire the design of our method, like the recurrent framework.
%a great progress, most methods are aimed at processing offline videos and are difficult to be applied in real-time rendering.

%Although the structure of the recurrent neural network can maintain good temporal stability, it also accumulates reconstruction errors, resulting in blurring results.% Our approach combines both structures and makes an effective trade-off. 

% However, in rendering, we can avoid this complex process and obtain the motion vector directly from the rendering pipeline. 

% \begin{figure*}
%   \centering
%   \includegraphics[width=0.95\linewidth]{figures/network10.pdf}
%    \caption{The overall architecture of our network. It mainly contains three modules: demodulation module, reliable warping module, and frame-recurrent super-resolution module. A low-resolution rendered image is first demodulated into a weighted-lighting component by the pre-integration BRDF in the demodulation module, and perform feature extraction together with other rendering auxiliary G-buffers. At the same time, the low-resolution historical frame information is input to extract reliable temporal features through a reliable warping module, and together with the features of the current frame, they are fed into the frame-recurrent super-resolution module for feature reconstruction. The reconstructed weighted-lighting component is re-modulated with the high-resolution pre-integration BRDF to obtain the final super-resolution result.}
%    \label{fig:network}
% \end{figure*}

\begin{figure*}
  \centering
  \includegraphics[width=0.95\linewidth]{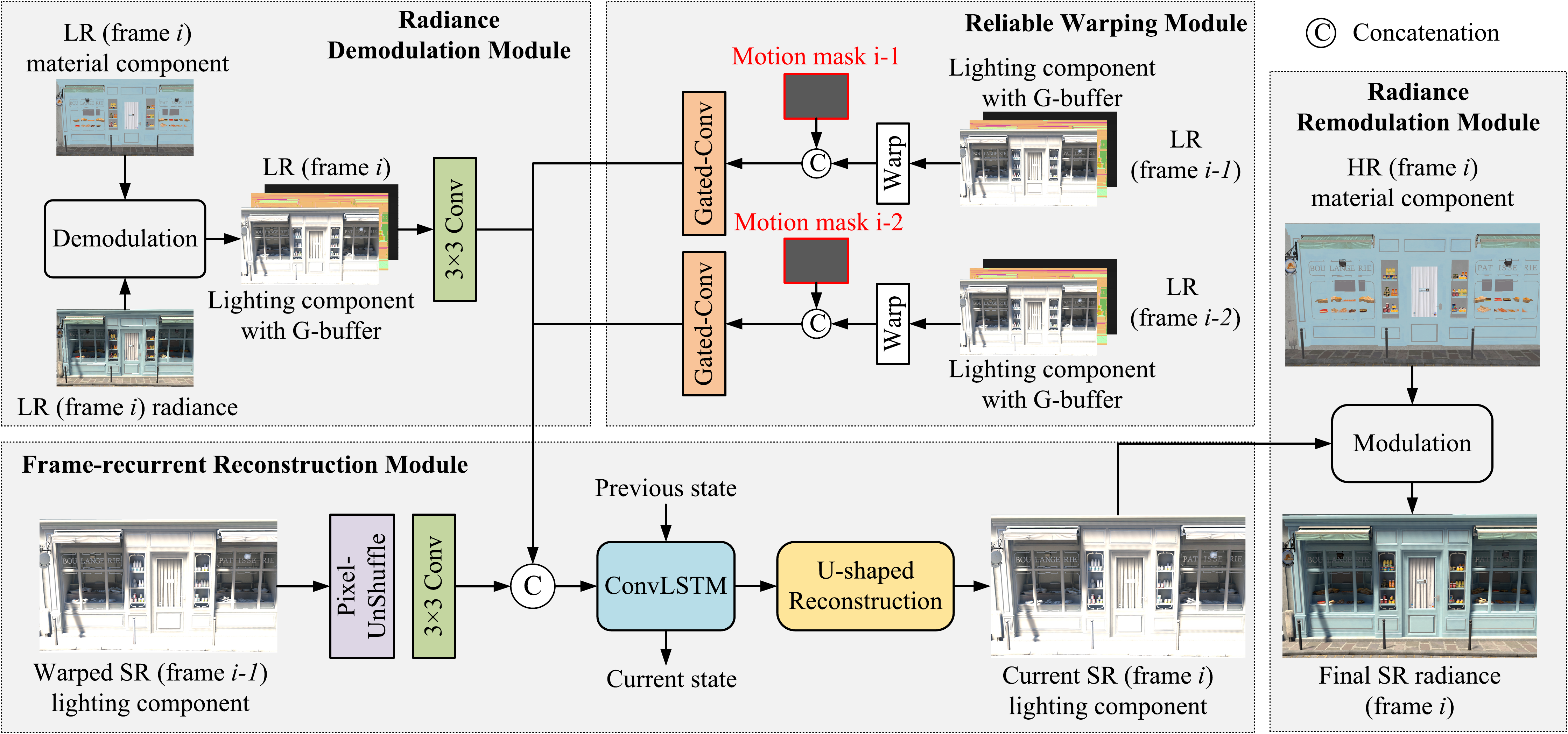}
   \caption{Our network includes four modules. \emph{Radiance demodulation:} together with the material component, the LR rendered image (radiance) is demodulated to a lighting component for spatial feature extraction. \emph{Reliable warping:} the warped lighting components of two previous frames and motion masks are fed into a gated convolution for temporal feature extraction. \emph{Frame-recurrent reconstruction:} features from the previously reconstructed SR lighting component and other features are fed into a ConvLSTM followed by a U-shaped module to reconstruct the SR lighting component, which is later re-modulated with the HR material component to obtain the SR image.}
   \label{fig:network}
\end{figure*}

% \subsection{Supersampling in Rendering}
\paragraph{Super-Resolution in Realtime Rendering.} 
Existing methods for RRSR mainly include the traditional ones~\cite{TAAU,FSR} and deep-learning-based approaches~\cite{DLSS,NSRR,SRPO,XeSS}. Temporal antialiasing upscaling (TAAU)~\cite{TAAU} simply uses temporal accumulation to perform SR. Edelsten et al.~\cite{DLSS} propose deep learning supersampling (DLSS), considering both temporal and spatial information. However, their method relies on NVIDIA's hardware platform, and there is no publicly available technical information. Unlike DLSS, FSR~\cite{FSR} does not rely on specific hardware platforms, and it generally achieves SR through upsampling and edge sharpening, with limited quality improvement, which is further improved by integrating TAAU, i.e., FSR 2.0. Intel's XeSS~\cite{XeSS} is also a deep-learning-based SR method for rendered images, which targets production, and only demo samples are available. 

Xiao et al.~\cite{NSRR} propose NSRR, using the UNet~\cite{unet} for SR reconstruction after zero-upsampling the features of multiple frames. However, NSRR suffers from ghosting artifacts in scenes with fast-moving objects or cameras. Gu et al.~\cite{SRPO} propose to extract edge features for SR interpolation but without considering temporal stability. Guo et al.~\cite{guo2022classifier} train two networks separately to classify pixels into different categories, and then predict weights for frame blending. Yang et al.~\cite{yang2023mnss} use an alternate sub-pixel sample pattern during rasterization to create a small SR model that can be run on mobile devices. Mercier et al.~\cite{mercier2023efficient} propose a lightweight recurrent network and a gaming dataset for RRSR. All of these methods can achieve real-time performance, but they are unable to restore complex texture details accurately. The concurrent work FuseSR~\cite{zhong2023fusesr} utilizes many HR auxiliary buffers and pre-integrated demodulation to improve the texture details, but its lack of design for temporal stability tends to cause flickering results.

Similar to the above methods, our method also targets real-time performance. However, we couple the rendering and SR to fully utilize the auxiliary buffer to compensate for the missing texture details, and employ a frame-recurrent design for temporal consistency between adjacent frames. Consequently, we achieve better reconstruction quality and temporal stability.

\section{Method}
% \label{sec:method}
% This section starts with a high level overview of our approach (Section~\ref{sec:overview}), and then goes into details about the various modules, including: \added{Radiance demodulation} (Section~\ref{sec:texture}), network architecture design (Section~\ref{sec:network}) and motion mask in occlusion (Section~\ref{sec:warp}).

%\subsection{Overview}
%\label{sec:overview}

Our method consists of four modules: a radiance demodulation module which extracts the spatial features coupled with a remodulation module (Section~\ref{sec:texture}), a reliable warping module (Section~\ref{sec:warp}) which extracts temporal features from the previous frame and a frame-recurrent reconstruction module (Section~\ref{sec:network}) which extracts features from the previously reconstructed images and performs the reconstruction as well as upsampling on all the concatenated features. The overview structure is shown in Figure~\ref{fig:network}.

\subsection{Radiance Demodulation}
\label{sec:texture}

Most existing methods perform SR on the rendered images (radiance). Besides this rendered image, existing works~\cite{NSRR, guo2022classifier, mercier2023efficient} also use G-buffer (e.g., depth) to aid the SR process, allowing geometry details reconstruction. However, the G-buffer can not assist the reconstruction of the detailed textures. 
% A well-known knowledge in rendering is that the radiance is a convolution of the lighting and materials, where the high-frequency texture details mainly come from the material part, and the lighting part tends to be smooth. The decoupling of the lighting and materials is called radiance demodulation, and it is widely used in denoising ~\cite{KPCN}.

A well-known knowledge in rendering is that the radiance is a convolution of the lighting and materials. The texture details mainly come from the material component, and the lighting component tends to be smoother than the radiance, as shown in the supplementary. The decoupling of the lighting and materials is called radiance demodulation. The other key observation is that the material component, even in HR, can be generated efficiently, since it does not need global light transport. Motivated by these insights, we introduce radiance demodulation into our SR pipeline. 
% We demodulate the radiance to the lighting and material component during rendering, and then perform SR on the lighting component. We also provide a HR image of the material component, which will be used for the re-modulation. 

The radiance for a diffuse material can be directly demodulated into a lighting component (called irradiance) and a material component (called albedo), which is widely used in denoising~\cite{KPCN, kontkanen2004irradiance, schied2017svgf}. However, this idea only works for diffuse materials, and it's not applicable for a view-dependent material (e.g., glossy material). Recently, Zhuang et al.~\cite{pre-in} partially separate a material component from the radiance image, significantly aiding the denoising task.

 % The next question is how to demodulate the radiance. The radiance for a diffuse material can be demodulated into a lighting component (called irradiance) and a material component (called albedo) directly. However, it is impossible to decouple the lighting and the material entirely for a view-dependent material (e.g., glossy material). Recently, Zhuang et al.~\cite{pre-in} partially separate a material component from the radiance image, significantly aiding the denoising task. 

Following Zhuang et al.~\cite{pre-in}, the rendering equation~\cite{Kajiya:1986:rendering_equation} is reformulated as:
\begin{eqnarray}
\label{eq:reform}
L(\omega_o) & = & \int L(\omega_i) \rho (\omega_i,\omega_o) \cos{\theta_i} \mathrm{d} \omega_i,
\\ & = & F_{\beta}(\omega_o) \cdot I(\omega_o),
\end{eqnarray}
where
\begin{eqnarray}
\label{eq:Fb}
F_{\beta}(\omega_o) & = & \int \rho (\omega_i,\omega_o) \cos{\theta_i} \mathrm{d} \omega_i
\\ I(\omega_o) & = & \frac{\int L(\omega_i) \rho (\omega_i,\omega_o) \cos{\theta_i} \mathrm{d} \omega_i}{F_{\beta}(\omega_o)}
\end{eqnarray}
$L(\omega_o)$ and $L(\omega_i)$ represent the radiance at the outgoing direction $\omega_o$ and the incident direction $\omega_i$, respectively. $\rho$ represents the bidirectional reflectance distribution function (BRDF) and $\theta_i$ is the angle between the incoming direction and the shading normal. $F_{\beta}$ and $I$ represent the material component and the lighting component, respectively.
% Although the lighting component includes partial material information, we believe this separation is already good enough for the super-resolution task.}

% \begin{eqnarray}
% \label{eq:reform}
% L(\omega_o) & = & \int L_i(\omega_i) \rho (\omega_i,\omega_o) \cos{\theta_i} \mathrm{d} \omega_i,
% \\ & = & F_{\beta} \cdot \frac{\int L_i(\omega_i) \rho (\omega_i,\omega_o) \cos{\theta_i} \mathrm{d} \omega_i}{F_{\beta}}, 
% \\ & = & F_{\beta} \cdot I,
% \end{eqnarray}
% where
% \begin{equation}
% \label{eq:Fb}
% F_{\beta} = \int \rho (\omega_i,\omega_o) \cos{\theta_i} \mathrm{d} \omega_i.
% \end{equation}
% $L_i(\omega_i)$ represents the incoming radiance at incoming direction $\omega_i$, $\rho$ represents the bidirectional reflectance distribution function (BRDF) and $\theta_i$ is the angle between the incoming direction and the shading normal. $F_{\beta}$ and $I$ represent the material component and the lighting component, respectively. Although the lighting component includes partial material information. But we believe this separation is already good enough for the super-resolution task.

Now, we fit demodulated components into our SR pipeline. The renderer provides an LR lighting component $I$ and an HR material component $F_{\beta}$. The SR is performed on $I$ with the neural network, and then the reconstructed $I$ is multiplied with the HR material component to get the final SR result of the radiance image. The details of $I$ and $F_{\beta}$ computation are shown in the supplementary material.

\subsection{Reliable Warping}
\label{sec:warp}

\begin{figure}[t]
  \centering
   \includegraphics[width=1.0\linewidth]{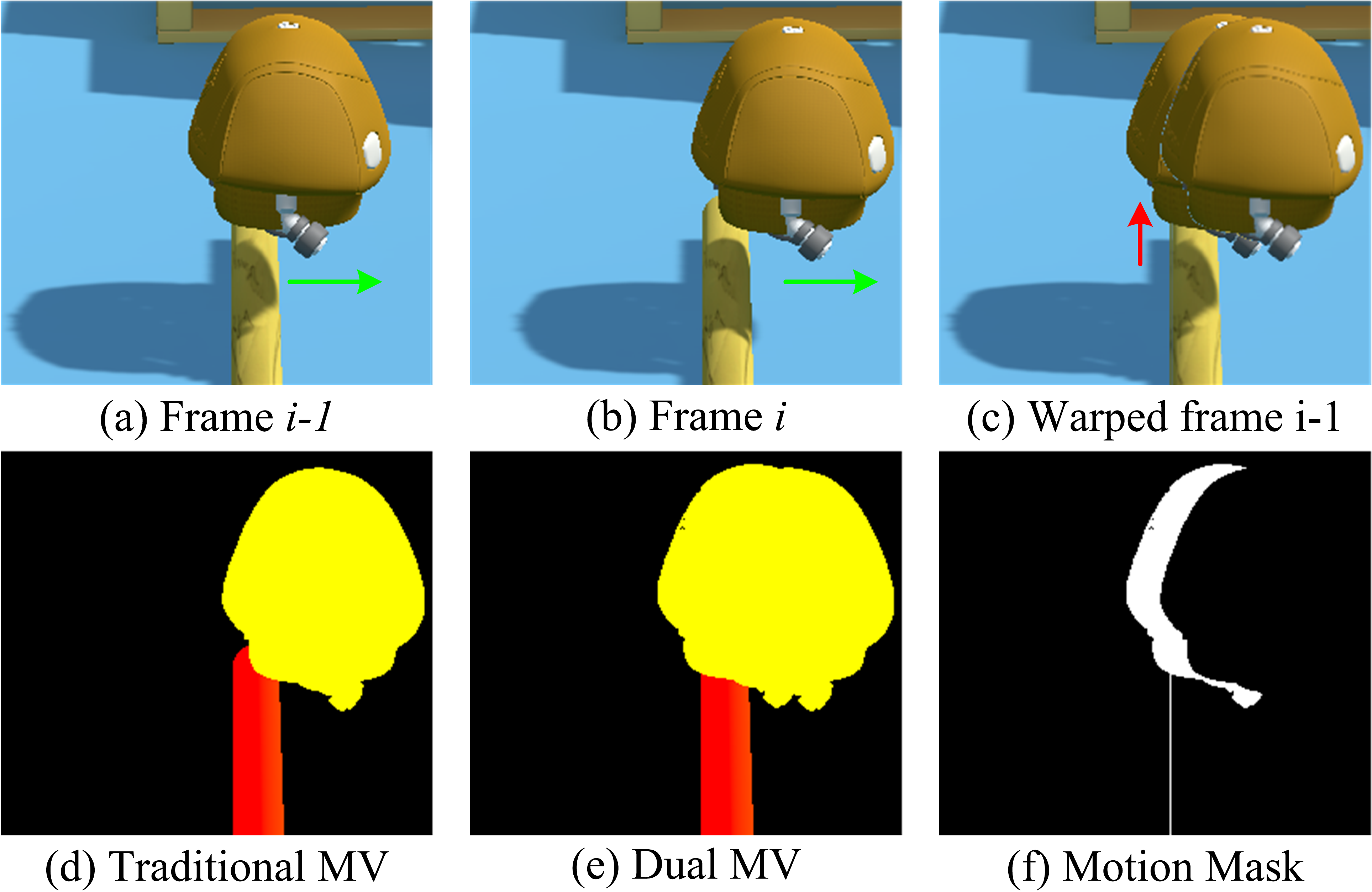}
   \caption{An example of the motion mask generation. Figures (a)-(b) represent the radiance of frame $i-1$ and frame $i$, respectively. The green arrows show the character's moving direction. Figure (c) is the warped result of frame $i-1$ using the traditional MV, where the occluded region has been warped incorrectly (red arrow). Figures (d)-(e) show the traditional MV and dual MV respectively. Figure (f) shows our motion mask, where the previously occluded region is marked.}
   \label{fig:mask}
   \vspace{-1.2em}
\end{figure}

Most existing VSR and RRSR methods warp the previous frames to the current frame, using optical flow or motion vector~\cite{MV}. For RRSR, the motion vector can be easily obtained during rendering. However, the motion vector becomes unreliable due to the occlusion of objects in dynamic scenes. As shown in Figure~\ref{fig:mask} (c), the regions (pointed by the red arrow) which are occluded in the previous frame but without occlusion in the current frame have been incorrectly warped into the current frame. These regions are called \emph{motion-unreliable regions}, leading to ghosting artifacts since the SR network is not aware of these regions. Hence, we need to explicitly point out the unreliable regions for the network to aid the SR.
%Previous work~\cite{NSRR} trimainly deal with incorrectly warping regions implicitly through the reweighting module, and there are no more clues for network to determine whether it is an incorrect area, resulting in an inability to accurately handle fast-moving objects. 

In this paper, we recognize these unreliable regions accurately and design a so-called \emph{motion mask} in our network with a gated convolution so that the motion-unreliable regions in the warped previous frames have small weights.

% \vspace{-1.2em}
\paragraph{Motion Mask.} We propose a novel yet simple idea to recognize the motion-unreliable regions. First, the traditional motion vector (TMV) can show the area of occluders (moving objects) in the current frame, as shown in Figure~\ref{fig:mask} (d). Second, a dual motion vector (DMV) proposed by Zeng et al.~\cite{TRMV} is able to recognize the area of occluders in the previous and the current frame together (more details can be found in supplementary), as shown in Figure~\ref{fig:mask} (e). To recognize the motion-unreliable region, we first subtract DMV with TMV and perform a binarization, which sets the pixel as 0 when its value is smaller than a threshold (0.1 in practice), otherwise set as 1. Then, we get a motion mask, which accurately marks the previously occluded regions, as shown in Figure~\ref{fig:mask} (f). 

\paragraph{Gated Convolution.} Then, we use the motion mask in our network with a gated convolution~\cite{gated-conv} to enable a learnable feature selection mechanism. The network can better identify unreliable regions and use less temporal information in these regions. Specifically, the warped LR lighting component and G-buffers are concatenated with the motion mask to form a reliable-aware input. Then a dynamic gating map is learned from the reliable-aware input through a convolution. The valid and invalid pixels in the reliable-aware features extracted from a convolution can be treated differently using this dynamic gating map. It is formulated as:
\begin{equation}
\begin{aligned} 
    X_{\mathrm{gate}}&=\mathrm{Conv}(W_{g},X_{\mathrm{in}}) ,\\
    X_{\mathrm{feat}}&=\mathrm{Conv}(W_{f},X_{\mathrm{in}}) ,\\
    X_{\mathrm{out}}&=\phi( X_{\mathrm{feat}}) \odot \sigma(X_{\mathrm{gate}}).
\end{aligned}
\end{equation}
where $X_\mathrm{in}$ is the reliable-aware input, $X_{\mathrm{gate}}$ is the dynamic gating map, $X_{\mathrm{feat}}$ is reliable-aware features, and $X_{\mathrm{out}}$ is the output feature map from the reliable warping module. $W_{g}$ and $W_{f}$ are two convolutional filters, $\odot$ denotes element-wise multiplication, $\phi$ and $\sigma$ indicate the LeakyReLU activation function and sigmoid function respectively.

%\textbf{Gated Convolution.} Instead of the traditional convolution, we adopt the gated convolution~\cite{gated-conv} which enables a learnable feature selection mechanism, using the motion mask so that the network can better identify the regions with incorrect reprojection and will pay less attention to these regions. Specifically, a dynamic weighting map is learned from the input features with mask information. Using this dynamic weighting map, valid and invalid pixels can be treated differently.

% \added{Beibei: we need to talk about these equation. Where is the mask? We have not talked about a input feature map yet. }
%The calculation is as follows:
%\begin{equation}
%    Gating=Conv(W_{g},X_{\mathrm{in}})
%\end{equation}
%\begin{equation}
%    Feature=Conv(W_{f},X_{\mathrm{in}})
%\end{equation}
%\begin{equation}
 %   X_{\mathrm{out}}=\phi(\mathrm{Feature}) \odot \sigma(\mathrm{Gating})
%\end{equation}
%where $X_{in}$ is the input feature map and $X_{\mathrm{out}}$ is the output feature map, $W_{g}$ and $W_{f}$ are two different convolutional filters, $\odot$ denotes element-wise multiplication, $\phi$ and $\sigma$ indicate the LeakyReLU activation function and sigmoid function respectively.

\subsection{Frame-Recurrent Super-resolution}
\label{sec:network}

Until now, we already have the features extracted from the current and previous frames, which can enable a super-resolution. However, we notice a temporal-unstable issue. Inspired by previous works~\cite{FRVSR, haris2019rcnn, FFCVSR, fuoli2019RLSP, RSDN, chan2022basicvsr++}, the recurrent neural networks can reduce flickers between frames and produce a relatively stable video by using the reconstructed image from the previous frame. Thus, we also integrate the recurrent framework into our network.

%We could concatenate them and perform a feature reconstruction and upsampling to get the final SR result. 
%this is kind of overview.
In our recurrent framework, we not only introduce the previously reconstructed image but also exploit a convolution LSTM (ConvLSTM)~\cite{conLSTM} structure for better spatial-temporal feature extraction. The ConvLSTM can effectively utilize the state of intermediate features to obtain more spatial-temporal correlation between adjacent frames, so as to improve the reconstruction quality. Later, we use a U-shaped reconstruction module for residual channel attention blocks~\cite{RCAN} connection to reduce network computation.

%this is the detail.
Now, we show the entire frame-recurrent reconstruction module. Firstly, we warp the reconstructed lighting component from the previous frame to the current frame, perform a pixel-unshuffle~\cite{Shi:2016:ESPCN} operation, which maps the HR input to LR space, and extract its features and concatenate the shallow features of the other two modules (radiance demodulation and reliable warping module). Then the concatenated features are fed into the ConvLSTM together with the states of the previous intermediate features, and then fed into a U-shaped reconstruction module to reconstruct the HR lighting component of the current frame. The structure of the reconstruction module can be found in the supplementary.

Our loss function consists of three parts: a smooth $L_{1}$ loss~\cite{smL1} ($L_{1}^\mathrm{s}$), a structural similarity loss~\cite{SSIM} ($L_\mathrm{SSIM}$) and a temporal loss ($L_t$). The temporal loss can better ensure the coherence of adjacent reconstructed frames. The final loss function is defined as:
\vspace{-1.2em}
%which are combined by weights, as shown in the formula~\ref{eq:loss}.
%In addition to considering the spatial loss of the result in our training process, we also introduce the temporal loss function in order to better ensure the temporal stability of the reconstructed results. Therefore, the loss function consists of three parts, namely Smooth $L_{1}$ loss~\cite{smL1}, SSIM loss~\cite{SSIM} and temporal loss, which are combined by weights, as shown in the formula~\ref{eq:loss}.
% \begin{equation}
% \begin{aligned}
% L_\mathrm{final}(\hat{I}_{i-1}^{\mathrm{SR}},I_{i}^{\mathrm{SR}},I_{i}^{\mathrm{HR}}) =& w_1 \cdot L_{1}^\mathrm{s}(I_{i}^{\mathrm{SR}},I_{i}^{\mathrm{HR}})\\+ w_2 \cdot (1- L_\mathrm{SSIM}(I_{i}^{\mathrm{SR}},I_{i}^{\mathrm{HR}}))+&w_3 \cdot L_t(\hat{I}_{i-1}^{\mathrm{SR}},I_{i}^{\mathrm{SR}}),
% \end{aligned}
% \label{eq:loss}
% \end{equation}
% where $L_t$ is defined as:
% \begin{equation}
% L_t(\hat{I}_{i-1}^{\mathrm{SR}},I_{i}^{\mathrm{SR}})=L_{1}^\mathrm{s}(M_{i-1}\odot\hat{I}_{i-1}^{\mathrm{SR}},M_{i-1}\odot I_{i}^{\mathrm{SR}}).  
% \label{eq:Lt}
% \end{equation}

\begin{equation}
\begin{aligned}
L_\mathrm{final}(\hat{I}_{i-1}^{\mathrm{SR}},I_{i}^{\mathrm{SR}},I_{i}^{\mathrm{HR}}) &= w_1 \cdot L_{1}^\mathrm{s}(I_{i}^{\mathrm{SR}},I_{i}^{\mathrm{HR}})\\&+ w_2 \cdot (1- L_\mathrm{SSIM}(I_{i}^{\mathrm{SR}},I_{i}^{\mathrm{HR}}))\\&+w_3 \cdot L_t(\hat{I}_{i-1}^{\mathrm{SR}},I_{i}^{\mathrm{SR}}),
\end{aligned}
\label{eq:loss}
\end{equation}
where $L_t$ is defined as:
\begin{equation}
L_t(\hat{I}_{i-1}^{\mathrm{SR}},I_{i}^{\mathrm{SR}})=L_{1}^\mathrm{s}(M_{i-1}\odot\hat{I}_{i-1}^{\mathrm{SR}},M_{i-1}\odot I_{i}^{\mathrm{SR}}).  
\label{eq:Lt}
\end{equation}

$\hat{I}_{i-1}^{\mathrm{SR}}$ represents the warped reconstructed lighting component of frame $i-1$. $I_{i}^{\mathrm{SR}}$ is the network output (lighting component of frame $i$) and $I_{i}^{\mathrm{HR}}$ is the HR reference for the lighting component of frame $i$. $M_{i-1}$ is the inversion value of the HR motion mask at frame $i-1$, which is generated by using the bilinear upsampling of the traditional and dual motion vectors. The weights $w_1$, $w_2$ and $w_3$ are set to 1:1:1.

%We find that ghost artifacts will appear when the $\omega_3$ is set to a large value, which may be caused by the transition dependence on the reconstructed results of the previous frame.

\section{Experiments}
\label{sec:exp}
% There are two ways to use our method, one is to train a network separately for each scene to obtain higher reconstruction quality; the other is to train in multiple different scenes together to generalize to new scenes. 
Unless specified in the experiment, the following experiments train a network separately on each scene for a better reconstruction quality. And by default the SR factor is set as 4 $\times$ 4, and the target resolution is 1920 $\times$ 1080. The best and second-best quality or performance is shown in \textbf{bold} and \underline{underlined}, respectively. All results are calculated on RGB-channel. More experimental details and results can be found in the supplementary.

\begin{table*}
  \small
  \centering
  \setlength\tabcolsep{2pt}
  % \caption{\revise{Comparison between our method and the other six methods with four error measurements on seven scenes.}}
  \caption{Quality and performance comparisons between our method and the other six methods on seven scenes.}
  \vspace{-0.8em}
  %Quality comparisons with 4 $\times$ 4 super-resolution. }
  \label{tab:qual}
  \begin{tabular}{ll|c|c|c|c|c|c|c}
    \hline
    &   & FRVSR & TecoGAN & NSRR & BasicVSR++ & TTVSR  & RVRT & Ours \\
    \hline
    \multirow{7}*{\rotatebox{90}{PSNR / SSIM }} 
    & Bistro      & 24.24 / 0.7648 & 23.02 / 0.7264 & 24.62 / 0.7975 & 25.31 / 0.8085 & 25.37 / 0.8108 & \underline{25.42} / \underline{0.8145} &{\textbf{26.43}} / {\textbf{0.8739}}\\
    & San\_M      & 27.57 / 0.8422 & 26.61 / 0.8141 & 27.52 / 0.8598 & 29.02 / 0.8752 & \underline{29.20} / \underline{0.8754} & 28.58 / 0.8608 &{\textbf{30.37}} / {\textbf{0.9426}}\\
    & Square      & 20.61 / 0.5880 & 19.32 / 0.5230 & 20.66 / 0.6009 & 21.13 / \underline{0.6276} & \underline{21.19} / 0.6253 & 20.95 / 0.6120 &{\textbf{21.58}} / {\textbf{0.7306}}\\
    & Bar         & 23.80 / 0.7800 & 23.48 / 0.7617 & 25.51 / 0.8445 & 26.20 / \underline{0.8513} & 26.02 / 0.8469 & \underline{26.22} / 0.8473 &{\textbf{27.16}} / {\textbf{0.9202}}\\
    & ZeroDay     & 21.53 / 0.7735 & 21.04 / 0.7624 & 21.82 / 0.7922 & 22.28 / 0.8125 & \underline{22.60} / \underline{0.8159} & 22.57 / 0.8142 &{\textbf{23.63}} / {\textbf{0.8680}}\\
    & Airplane    & 33.03 / 0.9450 & 32.27 / 0.9347 & 31.75 / 0.9429 & 33.77 / 0.9534 & 33.94 / \underline{0.9550} & \underline{33.98} / 0.9536 &{\textbf{34.09}} / {\textbf{0.9643}}\\
    & Pica        & 32.73 / 0.9621 & 31.34 / 0.9510 & 32.57 / 0.9630 & 35.19 / 0.9773 & 36.33 / 0.9818 & \textbf{37.09} / \underline{0.9821} &\underline{37.03} / {\textbf{0.9828}}\\
     \hline
    \multirow{7}*{\rotatebox{90}{$\downarrow$ LPIPS / VMAF}}
    & Bistro      & 0.326 / 35.97 & \underline{0.234} / 35.20 & 0.281 / 43.16 & 0.282 / 45.55 & 0.281 / 47.28 & 0.278 / \underline{48.87} &{\textbf{0.141}} / {\textbf{53.82}}\\
    & San\_M      & 0.276 / 46.21 & \underline{0.214} / 37.44 & 0.242 / 54.62 & 0.221 / 58.77 & 0.222 / \underline{59.66} & 0.244 / 54.13 &{\textbf{0.075}} / {\textbf{73.50}}\\
    & Square      & 0.443 / 17.57 & \underline{0.347} / 19.34 & 0.433 / 19.70 & 0.403 / 19.34 & 0.408 / \underline{23.26} & 0.418 / 21.65 &{\textbf{0.227}} / {\textbf{26.26}}\\
    & Bar         & 0.321 / 34.13 & \underline{0.279} / 31.15 & 0.318 / 34.66 & 0.311 / 39.44 & 0.314 / \underline{40.83} & 0.319 / 40.01 &{\textbf{0.087}} / {\textbf{55.22}}\\
    & ZeroDay     & 0.301 / 21.43 & \underline{0.282} / 19.85 & 0.291 / 27.54 & 0.285 / 35.61 & 0.278 / \underline{37.32} & 0.282 / 36.92 &{\textbf{0.154}} / {\textbf{53.38}}\\
    & Airplane    & 0.186 / 64.43 & 0.153 / 65.77 & 0.186 / 56.99 & 0.148 / 70.16 & \underline{0.142} / \textbf{71.46} & 0.144 / 70.85 &{\textbf{0.076}} / \underline{70.92}\\
    & Pica        & 0.078 / 66.67 & 0.054 / 63.09 & 0.064 / 67.31 & 0.046 / 78.92 & 0.039 / 83.02 & \underline{0.030} / {\textbf{85.92}} &{\textbf{0.029}} / \underline{85.43}\\   
     \hline\hline
    & Params (M)   &  2.59              &  2.59               & {\textbf{0.53}}  & 7.32    & 6.77    & 10.78  &  \underline{1.61}   \\
    & Runtime (ms) &  \underline{14.97} &  \underline{14.97}  & 23.94            & 98.69   & \textgreater 100  & \textgreater 100   & {\textbf{12.41}}    \\   
    \hline
    % & GFLOPs        &  285.97  &  285.97 & 304.87             & 631.32 & 1394.68 & xxxx   & {\textbf{145.36}}    \\
\end{tabular}
\vspace{-1.0em}
\end{table*}

\subsection{Datasets and Implementation}
\label{sec:dataset}

%\paragraph{Datasets.} 
% We use the Unity~\cite{unity} rendering engine to generate our dataset. We select a total of 5 scenes, namely Bistro~\cite{Bistro}, Square~\cite{ORCANVIDIAEmeraldSquare}, San Miguel (San\_M)~\cite{McGuire:2017:Data}, Airplane and PICA. Each of these scenes has a rapid-moving camera or object, which is common in video games. We generate multiple sequences for each scene, each containing 300 frames. A detailed number of sequences and the partition of training, validation and testing datasets as well as example images are shown in the supplementary. We use the Disney material model~\cite{burley2012disneyPBR} in our scenes and render the scenes with a deferred rendering pipeline. 

% Large datasets are necessary for training robust networks, so 
Our dataset consists of seven representative scenes rendered with Unity~\cite{unity} engine, covering typical challenging scenarios in real-time rendering: complex textures and geometries (e.g., Bistro~\cite{Bistro}, Square~\cite{ORCANVIDIAEmeraldSquare}, San\_M~\cite{McGuire:2017:Data}), glossy reflections (e.g., Bar~\cite{Bistro} and ZeroDay~\cite{ZeroDay}) and fast-moving objects (e.g., Airplane and Pica scene). For rendering, we use the Disney material model~\cite{burley2012disneyPBR} and the ray-traced global illumination method in Unity. Similar to previous work~\cite{NSRR}, we set up different fast-moving cameras in each scene to generate multiple sequences (100 frames each). Each sequence includes different objects and materials to enhance diversity. Then, we randomly divide the training, validation and testing datasets from these sequences.

% \begin{table}
%   \centering
%   \small
%   \caption{The sequence number of training, validation and testing dataset for each scene. Each sequence contains 100 frames.}
%   \label{tab:data}
%   \begin{tabular}{lcccc}
%     \toprule
%      Scene & {Training seq.} & {Validation seq.} & {Testing seq.}\\
%     \midrule
%      Bistro     &   54  &   6  &  12 \\
%      San\_M      &   54  &   6  &  6  \\
%      Square     &   42  &   6  &  6  \\
%      Bar        &   45  &   3  &  6  \\
%      ZeroDay    &   24  &   3  &  3  \\  
%      Pica       &   30  &   3  &  3  \\
%      Airplane   &   24  &   3  &  3  \\
%   \bottomrule
% \end{tabular}
% \end{table}

For the training data, we generate the LR lighting component, traditional motion vector, dual motion vector and G-buffers (normal, depth) as the inputs and then generate an HR lighting component as the ground truth (GT). For that, we first render it at 3840 $\times$ 2160 with 8$\times$ MSAA and then downscale the image to 1920 $\times$ 1080 with a 2 $\times$ 2 box filter to reduce aliasing. For the testing data, we generate the LR input data and the HR material component.
% Note that the HR material component can be generated quickly by the pre-computation.

% For the training data, we generate the radiance image, traditional motion vector, dual motion vector and G-buffers (normal, depth, roughness, metallic, albedo, the dot product of normal and view direction, and specular) for each frame with both low resolution and high resolution. To generate the GT images, we use a forward rendering pipeline, so that the MSAA can be used. We first render them at 3840 $\times$ 2160 with 8$\times$ MSAA and then downscale the images to 1920 $\times$ 1080 with 2 $\times$ 2 box filter to reduce aliasing. Different from the training data, the test data does not need the high-resolution radiance image. Note that these high-resolution auxiliary data can be generated in one shot in the deferred rendering pipeline very quickly.

We also perform data augmentation on the training dataset by randomly cropping different regions with size 96 $\times$ 96 at each LR frame, bringing the total training data to at least 5000 patches per scene. The test data keeps the original LR size without any cropping.

%\paragraph{Implementation.} 

% We use two LR previous frames. All the convolutional layer outputs are 32 channels, except for the RCABs in the U-shaped reconstruction module, whose output channel number is 64. 

Our network is implemented in the PyTorch~\cite{pytorch} framework with the Adam optimizer~\cite{kingma2014adam}. The total number of training epochs is 200, and the initial learning rate is set to $5e^{-4}$, which is halved every 100 epochs. The training samples are fed into the network in a batch size of 8. Training takes about 24 hours on a single NVIDIA RTX 3090 GPU per scene. 

% More details are provided in the supplementary.
%The initial learning rate is set to $5e^{-4}$, attenuated by half every 100 epochs, \added{and a total of 200 epochs of training.}

% All of our training and testing was done on a PC with an Intel 12900K CPU and an NVIDIA RTX 3090 GPU, using the PyTorch~\cite{pytorch} framework. 

 %The number of training epoch is 200, batch size is 8, and the initial learning rate is $5e^{-4}$, which is attenuated by half on 100 rounds. And we adopt Adam optimizer~\cite{kingma2014adam}.

\begin{figure*}
    \centering
    \includegraphics[width=1.0\textwidth]{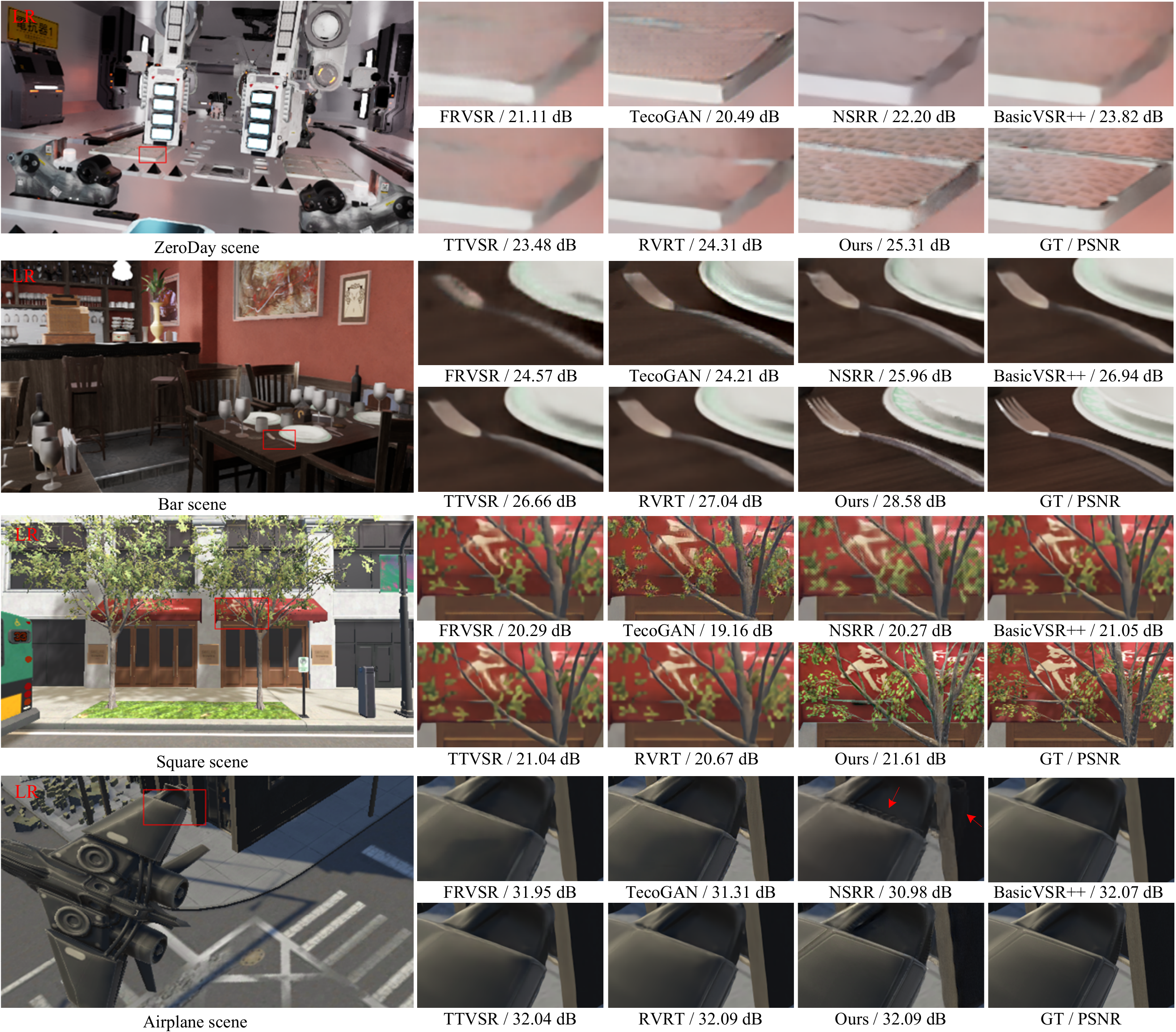}
    \caption{ Comparison among our method, FRVSR~\cite{FRVSR}, TecoGAN~\cite{tecoGAN}, NSRR~\cite{NSRR}, BasicVSR++~\cite{chan2022basicvsr++}, TTVSR~\cite{liu2022ttvsr} and RVRT~\cite{liang2022rvrt}. }%The PSNR metric is measured on the full image.}
    % \vspace{-1em}
    \label{fig:quality}
\end{figure*}

\begin{figure*}
    \vspace{+1em}
    \centering
    \includegraphics[width=1.0\textwidth]{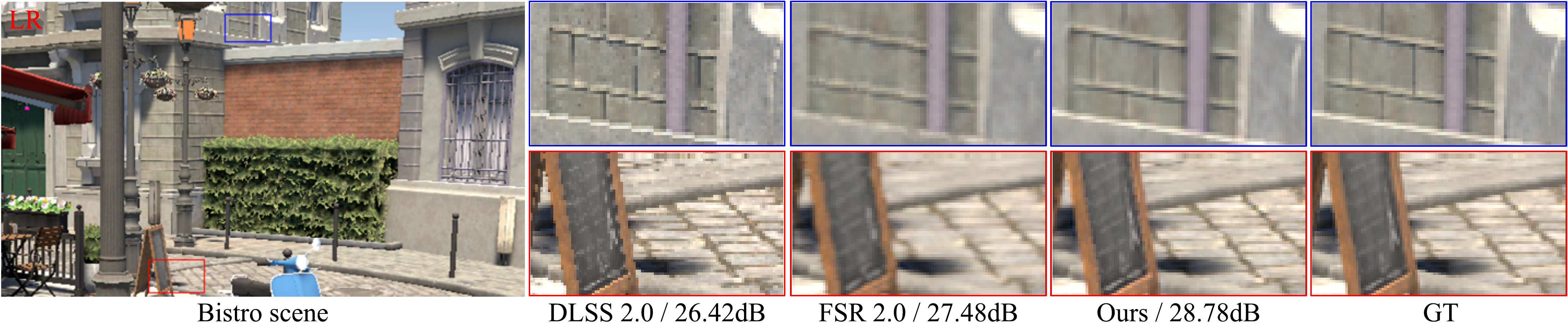}
    \caption{Comparison among our method, DLSS 2.0 ~\cite{DLSS} and FSR 2.0~\cite{FSR} on the Bistro scene. The target resolution is set as 1920 $\times$ 1080 and the SR factor is set as 2 $\times$ 2.}
    \label{fig:FSR_comp}
\end{figure*}

\begin{figure*}
    \centering
    \includegraphics[width=1.0\textwidth]{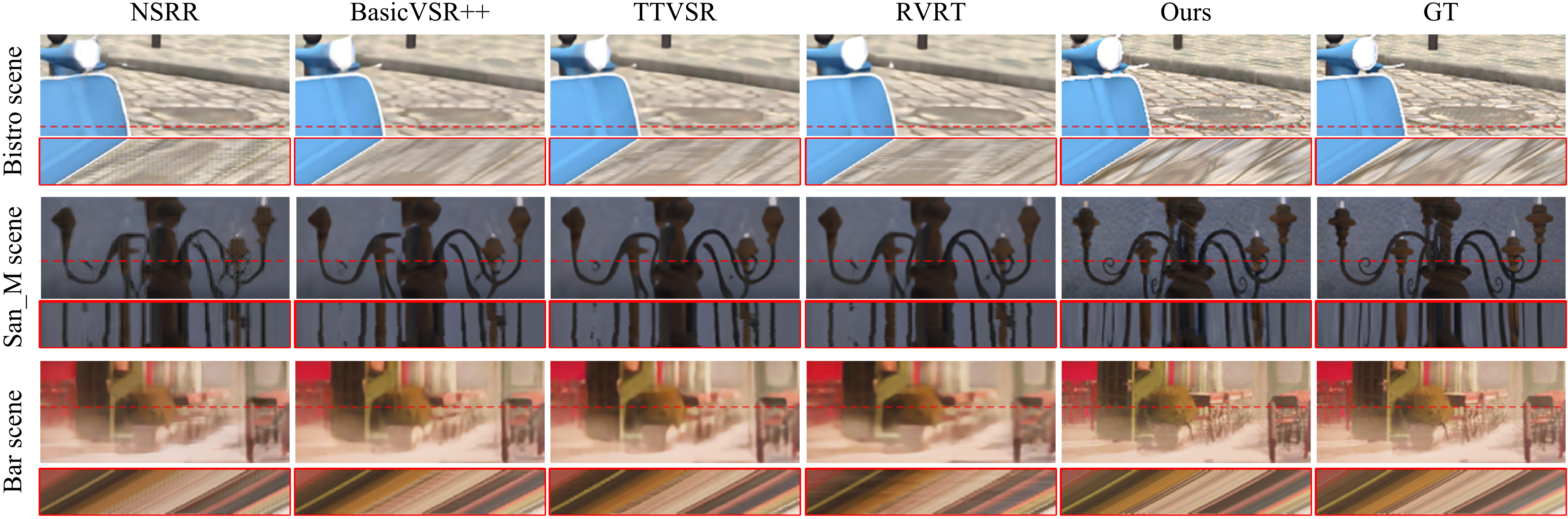}
    \caption{The EPI~\cite{bolles1987epipolar} comparison among NSRR, BasicVSR++, TTVSR, RVRT and our method on three scenes by plotting the transition of the dotted red horizontal scanline over time in the red box. The results which are sharper and closer to GT are better.}
    \label{fig:epi_main}
\end{figure*}

% \begin{figure}
%     \centering
%     \includegraphics[width=0.9\linewidth]{figures/ghost3.pdf}
%     \caption{ Comparison our method with NSRR~\cite{NSRR} on the Airplane scene. The red arrows point out the ghosting of NSRR. }
%     \label{fig:ghosting}
% \end{figure}

% \begin{table}
%   \small
%   \centering
%   \setlength\tabcolsep{2pt}
%   \caption{The sequence number of training, validation and testing dataset for each scene. Each sequence contains 100 frames.}
%   \label{tab:data}
%   \begin{tabular}{l|c|c|c|c|c|c|c}
%     \hline
%      Scene & Bistro & San\_M & Square & Bar & ZeroDay & Pica & Airplane\\     
%     \hline
%      Training   &  54  &  54  &  42  &  45  &  24  &  30  &  24 \\
%      Validation   &   6  &   6  &   6  &   6  &   3  &   3  &   3 \\
%      Testing    &  12  &   6  &   6  &   6  &   3  &   3  &   3 \\
%   \hline
% \end{tabular}
% \end{table}

\subsection{Quality Evaluation}
\label{sec:quality}
% To measure the quality of the reconstructed image, we use three quality metrics, peak signal-to-noise ratio (PSNR), structural similarity index (SSIM) \cite{SSIM} and video multi-method assessment fusion (VMAF) \cite{VMAF}, where VMAF considers temporal stability and perceptual quality.
We use four quality metrics to measure the quality of the reconstructed image: peak signal-to-noise ratio (PSNR), structural similarity index (SSIM) \cite{SSIM}, learned perceptual image patch similarity (LPIPS) \cite{zhang2018lpips} and video multi-method assessment fusion (VMAF) \cite{VMAF}, where VMAF considers temporal stability.

% We compare our method with five previous works:  FRVSR~\cite{FRVSR}, TecoGAN~\cite{tecoGAN}, BasicVSR++~\cite{chan2022basicvsr++}, and NSRR~\cite{NSRR}. NSRR and BasicVSR++ are state-of-the-art methods in the field of RRSR and VSR, respectively. Since NSRR does not have open-source code, we reproduce it and set the number of previous frames as 2, the same as our method. We retrain and test the other models on our dataset, using their released code. 
We first compare our method with six previous deep-learning-based VSR and RRSR methods:  FRVSR~\cite{FRVSR}, TecoGAN~\cite{tecoGAN}, BasicVSR++~\cite{chan2022basicvsr++}, TTVSR~\cite{liu2022ttvsr}, RVRT~\cite{liang2022rvrt} and NSRR~\cite{NSRR}. Since NSRR does not have open-source code, we reproduce it and set the number of previous frames as 2, the same as our method. We retrain and test the other models on our dataset using the released code. The reconstructed results of our method and the other six methods on five scenes are in Figures~\ref{fig:teaser} and \ref{fig:quality}. By comparison, our method can preserve more texture details, while all the other methods produce blurry results, such as the blackboard in the Bistro scene and painting in the Square scene, etc. Moreover, noticeable ghosting artifacts are shown in the results of NSRR (pointed by the red arrow), while our method is ghosting-free. We report the average quality metrics across all test data on seven scenes in Table~\ref{tab:qual}. Our method outperforms other methods on almost all scenes.

% We compare the reconstructed results among our method and other methods on seven scenes in Figures~\ref{fig:teaser} and \ref{fig:quality}. By comparison, our method can preserve more texture details, while all the other methods produce blurry results, such as the blackboard in the Bistro scene and painting in the Square scene, etc. Moreover, noticeable ghosting artifacts are shown in the results of NSRR (pointed by the red arrow in Figures~\ref{fig:teaser} and~\ref{fig:quality}), while our method is ghosting-free.

% We also report the quality metrics in Table~\ref{tab:qual}. Our method outperforms the other methods on all the metrics, and shows significant improvements.
We also compare our method with DLSS 2.0~\cite{DLSS} and FSR 2.0~\cite{FSR} in Table~\ref{tab:fsr} and Figure~\ref{fig:FSR_comp}. The SR factor is set as 2 $\times$ 2, since they officially do not support 4 $\times$ 4. By comparison, DLSS 2.0 has severe aliasing, and FSR 2.0 suffers from over-blur. Our method preserves more details while anti-aliasing, and outperforms them both qualitatively and quantitatively.% We believe our method will have greater advantages at higher SR factors with the help of the radiance demodulation module.}

 %We think this is because there are lots of indirect reflections in the scene, which cannot benefit from the material component and G-buffer inputs.}

We adopt the epipolar plane image (EPI)~\cite{bolles1987epipolar} to evaluate the temporal consistency visually by plotting the transition of the dotted red horizontal scanlines over time, as shown in Figure~\ref{fig:epi_main}. By comparison, our results look sharper and are the closest to the ground truth, while the other methods exhibit blurred results and flickering artifacts, demonstrating that our method is more temporal consistent than other methods.

\subsection{Performance Measurement}
\label{sec:perform}

% We report the parameter count, computation quantity (FLOPs) and running time of our network and four neural networks in Table~\ref{tab:perf}. We use Nvidia TensorRT~\cite{tensorRT} with 16-bit precision for acceleration on all the models to measure the inference time cost. Our computation and running time are much lower than other methods. Regarding the parameter count, NSRR has the fewest, but has a higher computational cost than ours. The main reason is that their upsampling is performed before reconstruction, leading to a larger feature size and higher computational cost. On the contrary, our method performs upsampling after the reconstruction, significantly reducing the computation.

We compare our method and others in terms of parameter count and running time at the bottom of Table~\ref{tab:qual}. We use Nvidia TensorRT~\cite{tensorRT} with 16-bit precision for acceleration on all models. Our running time are much lower than other methods. NSRR has the fewest parameters but with a longer running time than ours. The main reason is that its upsampling is performed before reconstruction, leading to a larger feature size and higher computational cost, while our method performs upsampling after the reconstruction, significantly reducing the computation.

We further analyze the runtime cost of our model in Table~\ref{tab:break}. The total time cost for our method is about 14.0 ms, which meets the real-time requirements and can be further shortened with some hardware acceleration. Note that generating the HR material component only costs 0.8 ms, which is negligible.

% Our SR method has improved the efficiency of real-time renderings significantly. For example, in the ZeroDay scene, rendering HR radiance directly with ray-traced GI costs about 89.6ms, while our method only costs 29.3ms, including 15.3ms for the LR lighting component rendering and 14.0ms for super-resolution. This leads to an over 3$\times$ performance improvement while maintaining high-fidelity results.

Our SR method has improved the efficiency of real-time renderings significantly. For example, in the ZeroDay scene, rendering HR radiance directly with ray-traced GI costs about 89.6ms. In contrast, it only costs 29.3ms using our method, including 15.3ms for the LR lighting component rendering and 14.0ms for the SR process. This leads to an over 3$\times$ performance improvement while maintaining high-fidelity results.

\subsection{Generalization Ability}
\label{sec:general}

Besides training each scene individually, our method can also be trained on several scenes and generalized to unseen scenes, trading the quality for generalization. To demonstrate the generalization ability of our model, we compare it with FRVSR, TecoGAN and NSRR in Table~\ref{tab:general} and the supplementary. We randomly select 120 sequences (12,000 frames in total) from five scenes, excluding the Bistro and Bar scenes, retrain all the methods and test them on the Bistro and Bar scenes. By comparison, our method produces higher quality than the other three methods, which indicates that our method has a generalization ability.

%This indicates that our method can generalize across scenes with different appearances.}

% \added{However, including the test scenes into training datasets seems to always improve the quality. We think this is because our method essentially learns the light energy distribution in the scene, and for different scenes, the complex light propagation is entirely different, meaning that the energy may be infinite or infinitesimal, which causes certain difficulties in generalizing the new scene.}

% \begin{table}
%   \centering
%   \small
%   \caption{Ablation experiment for the radiance demodulation and G-buffer inputs on the Bistro scene with target resolution set as 1080p and upsampling ratio set as 4 $\times$ 4.}
%   \label{tab:demod}
%   \begin{tabular}{lcccc}
%     \hline
%      Demodulation & G-buffer & PSNR(dB) & SSIM \\
%      \hline
%     \makecell[c]{$\times$}     & $\checkmark$     & 24.7771 & 0.80492 \\
%     \hline
%     \makecell[c]{$\checkmark$} & $\times$         & 25.8987 & 0.86546 \\
%     \hline
%     \makecell[c]{$\checkmark$} & $\checkmark$     & 26.4304 & 0.87393 \\
%   \hline
% \end{tabular}
% \end{table}

\subsection{Ablation Studies}
\label{sec:ablation}

% \paragraph{Radiance Demodulation.} To validate the impact of the radiance demodulation module, we compare the reconstructed results between our method (with radiance demodulation) and our method (w/o radiance demodulation) in Figure~\ref{fig:gbuffer}. The radiance demodulation preserves much more details on the blackboard and significantly improves the reconstruction quality (1.65dB in PSNR).

\begin{figure}[t]
  \centering
   \includegraphics[width=1.0\linewidth]{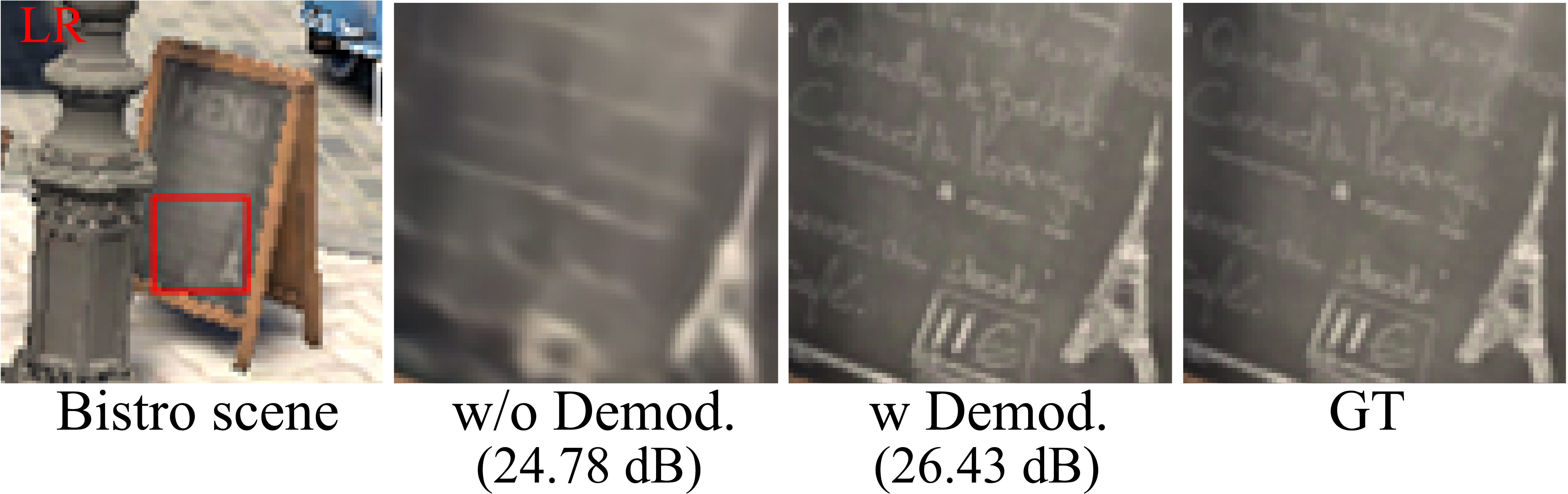}
   \caption{ Ablation study of the radiance demodulation. }%Below the cropped image is the PSNR corresponding to the whole image.
   \label{fig:gbuffer}
% \vspace{-1.2em}
\end{figure}

\paragraph{Radiance Demodulation.} To validate the impact of the radiance demodulation module, we provide a comparison in Figure~\ref{fig:gbuffer}. From the results, the radiance demodulation preserves more details on the blackboard and significantly improves the reconstruction quality (1.65dB in PSNR).

% Note that the radiance demodulation idea can also benefit other RRSR methods. We input the same G-buffer for NSRR as ours, and then further introduce the radiance demodulation. Then, we compare NSRR and two modified versions of NSRR with our method in Table~\ref{tab:comp_nsrr}. Introducing radiance demodulation improves moderately (0.67dB in PNSR) compared to G-buffer only. However, it still has a lower (0.98dB in PSNR) quality than ours. This also indicates that our method's specifically tailored network design plays an important role in our quality improvement.

Both G-buffer and radiance demodulation can benefit other methods (e.g., NSRR and BasicVSR++). We compare these two methods with two modified versions against our method in Table~\ref{tab:comp_nsrr}. Note that we only use the forward modules in BasicVSR++ (i.e., BVSR++\_Unidir.) for a fair comparison, since the bidirectional temporal feature propagation is infeasible in the real-time rendering applications. Introducing radiance demodulation improves moderately for NSRR (0.67dB in PSNR). However, it still has a lower (0.98dB in PSNR) quality than ours. BVSR++\_Unidir can achieve comparable quality (26.50dB vs. 26.43dB in PSNR) with ours, at the cost of 3$\times$ parameters count and 6$\times$ running time. It indicates that our method's lightweight network is tailored for real-time rendering.

\begin{table}
  \small
  \centering
  \caption{Comparison among our method, DLSS 2.0 and FSR 2.0 on the Bistro scene. The SR factor is set as 2$\times$2.}
  \label{tab:fsr}
  \begin{tabular}{l|c|c|c}
    \hline
     & PSNR(dB) & SSIM & LPIPS $\downarrow$   \\
     \hline
     DLSS 2.0 & 28.35 & 0.9104 & 0.136  \\
     FSR 2.0 & 28.90 & 0.9117  & 0.137   \\
     Ours   & \textbf{30.11} & \textbf{0.9405}  & \textbf{0.080}   \\
  \hline
\end{tabular}
\end{table}

\begin{table}
  \small
  \centering
  \caption{Time cost breakdown of our method. \textit{HR\_M} and \textit{LR\_G} represent the generation of HR material component and LR G-buffer respectively, and the \textit{Inference} means network inference.}
  \label{tab:break}
  \begin{tabular}{l|c|c|c|c|c}
    \hline
     & HR\_M & LR\_G & Warping & Inference & Total \\
     \hline
     {Time (ms)} & 0.84 & 0.35  & 0.43 & 12.41  & 14.03 \\
  \hline
\end{tabular}
\end{table}

\begin{table}
  \small
  \caption{Generalization ability comparisons with other three methods on Bistro and Bar scenes.}
  \label{tab:general}
  \begin{tabular}{ll|c|c|c|c}
    \hline
    & Scene  & FRVSR & TecoGAN & NSRR & Ours \\
    \hline
    \multirow{2}*{\tabincell{c}{PSNR\\(dB)}} 
    & Bistro      & 22.84 & 22.24 & 22.81 & {\textbf{23.29}} \\
    & Bar         & 22.13 & 22.02 & 22.07 & {\textbf{22.16}} \\
     \hline
    \multirow{2}*{{SSIM}}   
    & Bistro      & 0.7232 & 0.6964 & 0.7267 & {\textbf{0.8021}}\\
    & Bar         & 0.7488 & 0.7278 & 0.7352 & \textbf{{0.7836}}\\
  \hline
\end{tabular}
\end{table}

\textbf{Recurrent Framework and Temporal Loss.}
We study the impacts of our recurrent framework and temporal loss in Table~\ref{tab:temporal}. To measure the temporal stability, we introduce a warping error metric~\cite{lai2018warpError}, which is computed based on the flow warping error between adjacent frames. Combining the frame-recurrent framework and the temporal loss produces the best results. With frame-recurrent structure only, the PSNR and SSIM are better than the naive network, while the warping error is worse. Thus, the frame-recurrent framework improves the reconstruction quality, and the temporal loss improves the temporal stability.

%Our method adopts a frame-recurrent framework and uses a temporal loss function during training. In order to prove that the framework can play an important role in temporal stability, we introduce the warping error metric proposed in~\cite{lai2018warpError} to evaluate temporal stability, which is calculated based on the flow warping error between two frames. The results of ablation experiments are shown in Table~\ref{tab:temporal}. Note that the ConvLSTM~\cite{conLSTM} structure can only be used in recurrent neural networks. It can be seen from the table that the optimal results can be obtained when the frame-recurrent framework and the temporal loss are adopted simultaneously. In addition, when only frame-recurrent structure is introduced without temporal loss, the PSNR and SSIM will be better than the original network, but the warping error is worse. We believe that this is because the network only uses the previous reconstructed result to improve the current image quality in the training process, without considering the temporal stability.

\begin{table}
  \centering
  \small
  \setlength\tabcolsep{2pt}
  \caption{The impact of the radiance demodulation for NSRR and unidirectional BasicVSR++ (BVSR++\_Unidir.) on the Bistro scene. The quality metric is PSNR (dB).}
  \label{tab:comp_nsrr}
  \begin{tabular}{lcccc}
    \hline
                        &   +GBuffer  & +Demod.  &  Params (M) & Time (ms)  \\
    \hline
    NSRR                &   24.78   &   25.45    &  \textbf{0.54}   &  \underline{24.35}  \\
    BVSR++\_Unidir.     &   25.03   &   \textbf{26.50}    &  4.87   &  75.21  \\
    Ours                &   24.78   &   \underline{26.43}    &  \underline{1.61}   &  \textbf{12.41}  \\
  \hline
\end{tabular}
\end{table}

\begin{table}
  \centering
  \small
  \caption{Ablation experiment for the recurrent framework and temporal loss on the Bistro scene.}
  \label{tab:temporal}
  \begin{tabular}{lcccc}
    \hline
                        &  (A)    &  (B)            & Ours    \\
     \hline
    Recurrent Framework &         &  $\checkmark$   & $\checkmark$ \\
    Temporal Loss       &         &                 & $\checkmark$ \\
    \hline
    Warping Error $\downarrow$    &   3.5777   &  3.9011    &  2.9315    \\
    PSNR(dB)                      &   26.05    &  26.20      &  26.43    \\
    SSIM                          &   0.8662   &  0.8711    &  0.8739    \\
  \hline
\end{tabular}
\end{table}

\begin{figure}[t]
  \centering
   \includegraphics[width=1.0\linewidth]{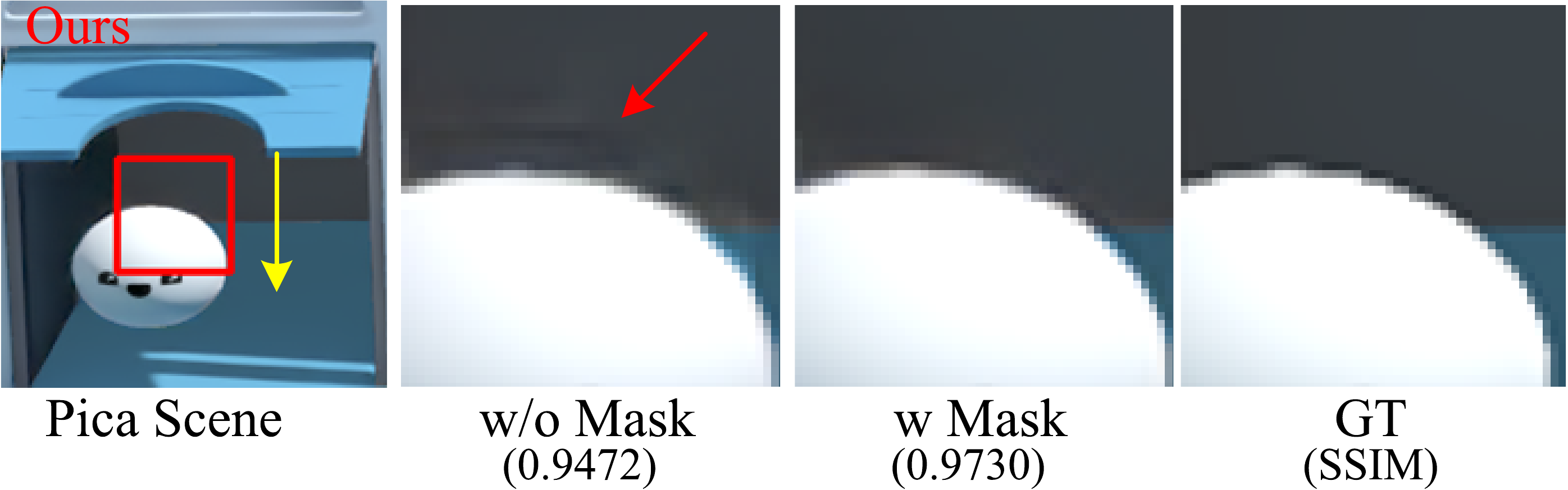}
   \caption{Ablation experiment for the motion mask (Mask). The yellow arrow indicates the direction of the ball's movement, the red arrow points out the ball's white ghosting. }%The SSIM is performed under the cropped image.}
   \label{fig:ghost_Ab}
   % \vspace{-1.2em}
\end{figure}

\textbf{Motion Mask.}
The impact of the motion mask is shown in Figure~\ref{fig:ghost_Ab}. Without the motion mask, a translucent ghosting of the ball can be seen from the red arrow, while the ghosting artifacts disappear after using the motion mask.

\section{Conclusion}
\label{sec:conclude}

In this paper, we have presented a novel lightweight super-resolution method for real-time rendering. By introducing radiance demodulation into super-resolution, the reconstructed quality is improved significantly. We also proposed a new approach to detect the motion-unreliable region, which serves as a mask to aid reconstruction and reduces the ghosting artifacts. Furthermore, a new frame-recurrent-based neural network is utilized to improve temporal stability while ensuring the reconstruction quality. Our method outperforms the existing state-of-the-art methods by a large margin. We believe it will benefit many applications like the interactive design or even video games, with further improvements in both quality and performance.

\section{Acknowledgments}
We thank the reviewers and Jin Xie for the valuable comments. This work has been partially supported by the National Natural Science Foundation of China under grants No.62272275 and No.62172220.

% In this paper, we have presented a novel lightweight super-resolution method for real-time rendering. By introducing radiance demodulation into super-resolution, the reconstructed quality is improved significantly. We also proposed a new approach to detect the motion-unreliable region, which serves as a mask to aid reconstruction and reduces the ghosting artifacts. Furthermore, a new frame-recurrent-based neural network is utilized to improve temporal stability while ensuring the reconstruction quality. Our method outperforms the existing state-of-the-art methods by a large margin. We believe that it will benefit many applications, like the interactive design or even video games, with further improvements in both quality and performance.

% In this work, we present a novel lightweight super-resolution method for real-time rendering. By introducing radiance demodulation into super-resolution, the reconstructed quality is improved significantly. We also proposed a new approach to detect the inaccurate motion vector region, which serves as a mask to aid reconstruction and reduce the ghosting artifacts. Furthermore, a new frame-recurrent based neural network are utilized to keep the temporal stability while ensuring the reconstruction quality. Our method outperforms existing state of the arts by a large margin and we believe it will benefit many applications with further improvements in the both quality and performance.

%%%%%%%%% REFERENCES
{\small
\bibliographystyle{ieee_fullname}
\bibliography{ref}
}

\end{document}

\title{Supplemental materials: Neural Super-Resolution for Real-time Rendering with Radiance Demodulation}

% \and
\maketitle

%%%%%%%%% BODY TEXT
\newcommand{\tabincell}[2]{\begin{tabular}{@{}#1@{}}#2\end{tabular}}

\section{Details of Radiance Demodulation}
\label{sec:sup_demod}

\begin{figure}[t]
  \centering
   \includegraphics[width=1.0\linewidth]{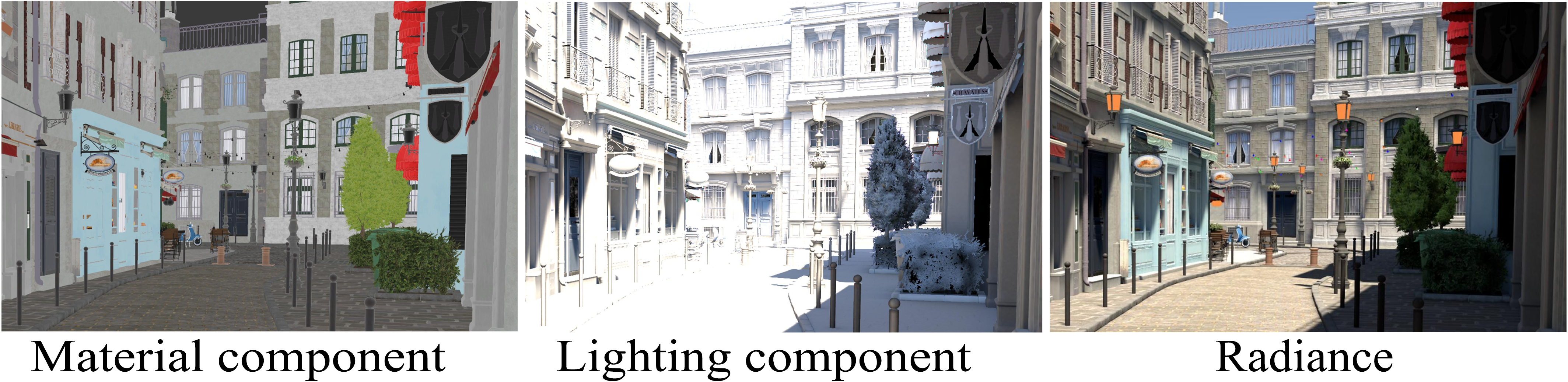}
   \caption{Visualization of the material component, lighting component and the radiance image. The texture details are shown on the material component, and the lighting component is much smoother than the radiance.}
   % \caption{Visualization of the material component, lighting component and the radiance image.}
 %  $F_\beta$, $L_\omega$ and $L(\omega_o)$. From left to right, the pre-integration BRDF, the weighted-lighting component and the rendered image (radiance), are shown respectively.}
   \label{fig:decomposition}
\end{figure}

In this section, we first introduce the specific bidirectional reflectance distribution function (BRDF) used in the implementation and then go into detail about the pre-computation of the material component $F_{\beta}$. One example of radiance demodulation is shown in Figure~\ref{fig:decomposition}.

\paragraph{Bidirectional Reflectance Distribution Function. }
We use Disney physically-based material model~\cite{karis:2013:shading} as our bidirectional reflectance distribution function (BRDF), which is widely used nowadays. And we can split the BRDF into diffuse and specular terms in real-time rendering. The formula is as follows:
\begin{align}
\label{eq:ds}
    \rho (\omega_i,\omega_o) & = \rho_{\mathrm{diff}}(\omega_i,\omega_o)+\rho_{\mathrm{spec}}(\omega_i,\omega_o),
\end{align}
where $\omega_i$ and $\omega_o$ represent the incoming direction and outcoming direction, respectively. $\rho_{\mathrm{diff}}$ and $\rho_{\mathrm{spec}}$ represent the diffuse term and specular term of the BRDF.

For the diffuse term, we directly use the Lambertian model, the formula is as follows:
\begin{align}
    \rho_{\mathrm{diff}}(\omega_i,\omega_o) & = k_{\mathrm{d}} \frac{c}{\pi}, \\
    & = (1-m)\frac{c}{\pi},
\end{align}
where $k_{\mathrm{d}}$ represents the diffuse ratio which can be calculated by metallic value $m$, and $c$ is the albedo of the object.

For the specular term, we use the Cook-Torrance~\cite{cook1982reflectance} microfacet specular shading model. The general formula is as follows:
\begin{align}
    \rho_{\mathrm{spec}}(\omega_i,\omega_o) & = \frac{D(\omega_h)F(\omega_o, \omega_h)G(\omega_i,\omega_o)}{4(\omega_o \cdot \omega_h)(\omega_i \cdot \omega_h)}, 
\end{align}
where $\omega_h$ represents the half vector between $\omega_i$ and $\omega_o$. $D(\omega_h)$ is the normal distribution function (NDF), $F(\omega_o, \omega_h)$ is the Fresnel term and $G(\omega_i,\omega_o)$ is the shadowing-masking function.

We use the GGX/Trowbridge-Reitz model~\cite{trowbridge1975average} as our normal distribution function:
\begin{equation} 
  D(\omega_h)=\frac{\alpha^2}{\pi((\omega_n \cdot \omega_h)^2(\alpha^2-1)+1)^2},
\end{equation}

where $\alpha$ and $\omega_n$ represent the roughness and normal of the object surface, respectively.

For the Fresnel term, we use Schlick's approximation~\cite{schlick1994inexpensive}:
\begin{equation} 
   F(\omega_o, \omega_h)=F_0+(1-F_0)(1-(\omega_o\cdot\omega_h))
\end{equation}
and
\begin{equation}
    F_0 = \mathrm{lerp}(0.04, c, m),
\end{equation}
where $F_0$ is the specular reflectance at normal incidence, which can be obtained by linear interpolation using the metallic value $m$ from plastic Fresnel coefficient (0.04) to albedo $c$.

% Finally, we use the Schlick model~\cite{schlick1994inexpensive} as our geometric function $G(\omega_i,\omega_o)$, and take into account both geometric shadowing and masking through the Smith method~\cite{walter2007microfacet}, the formula is as follows:
Finally, we use Smith method~\cite{walter2007microfacet} and Schlick model~\cite{schlick1994inexpensive} to formulate our shadowing-masking function:
\begin{equation} 
   G(\omega_i, \omega_o)=G_1(\omega_i)G_1(\omega_o),
\end{equation}
where 
\begin{align} 
   G_1(\omega_o) & = \frac{\omega_n \cdot \omega_o}{(\omega_n \cdot \omega_o)(1-k)+k}, \\
   k & = \frac{(\alpha + 1)^2}{8}. 
\end{align}

\paragraph{Pre-computation.} Following Zhuang et al.~\cite{pre-in} and the equation~\ref{eq:ds}, the material components $F_{\beta}$ in the radiance demodulation module can be rewritten as

\begin{align}
F_{\beta}(\omega_o) & = \int \rho (\omega_i,\omega_o) \cos{\theta_i} \mathrm{d} \omega_i,
\\ & = \int \rho_{\mathrm{diff}}(\omega_i,\omega_o)\cos{\theta_i}\mathrm{d} \omega_i \notag \\ 
& + \int \rho_{\mathrm{spec}}(\omega_i,\omega_o)\cos{\theta_i}\mathrm{d} \omega_i,
\end{align}    
where $\theta_i$ is the angle between the incoming direction and the shading normal. We split the integral into two terms, the diffuse term and the specular term.

The integral of the diffuse term can be calculated directly:
\begin{align}
\int \rho_{\mathrm{diff}}(\omega_i,\omega_o)\cos{\theta_i}\mathrm{d} \omega_i & = k_\mathrm{d} c, \\
& = (1 - m) c.
\end{align}

\begin{figure}
    \centering
    \includegraphics[width=0.5\linewidth]{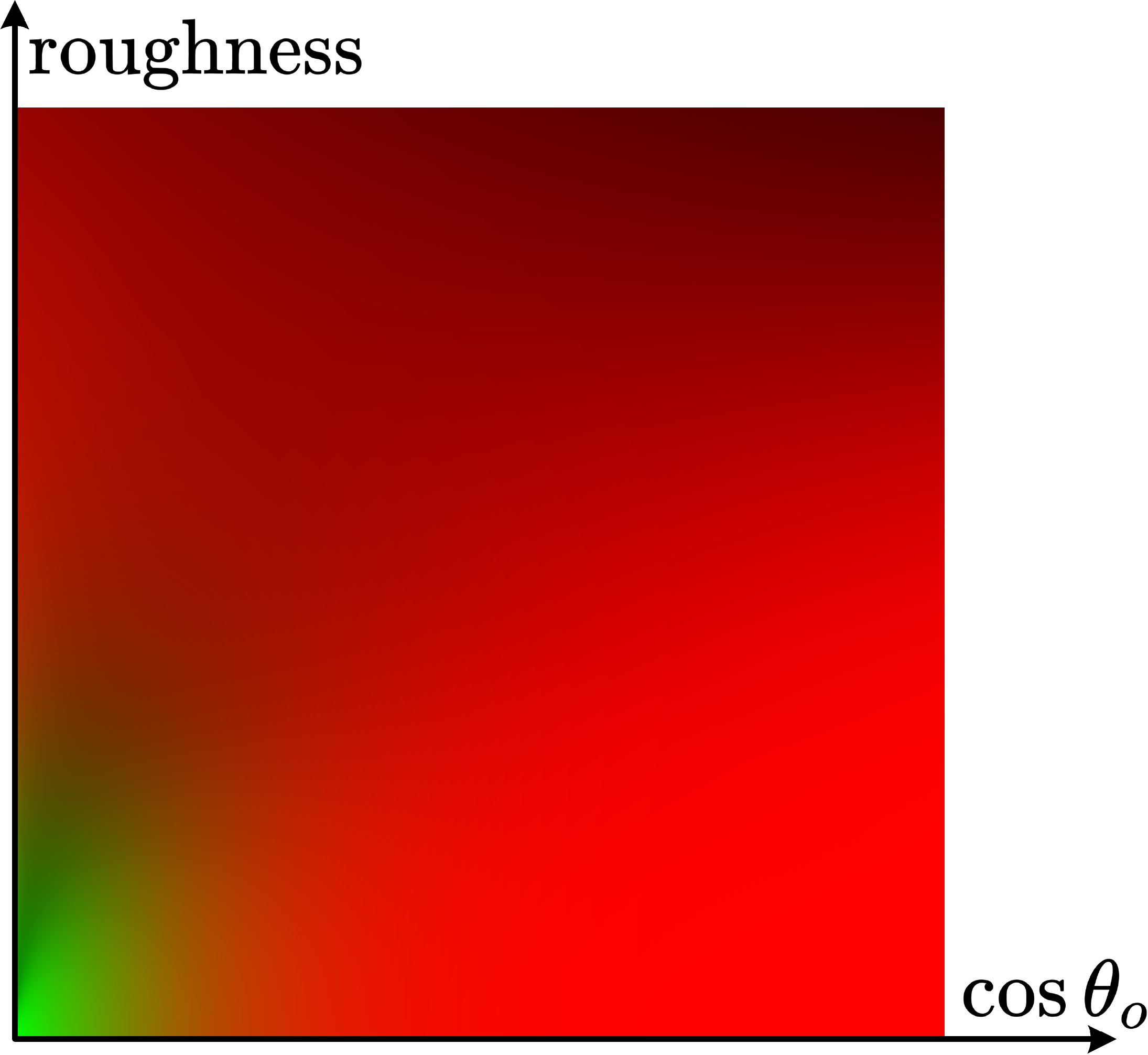}
    \caption{Pre-computation lookup table. The horizontal axis is $\cos{\theta_o}$, and the vertical axis is roughness $\alpha$. The first and second channels are the scale (denoted by $A$) and the bias (denoted by $B$) to $F_0$, respectively.}
    \label{fig:lut}
\end{figure}

Because the integral of the specular term is angular-dependent, we need to pre-compute this part to convert it into a simple function. However, if we directly pre-compute the entire integral, we need to store many parameters. Inspired by Karis et al.~\cite{karis:2013:shading}, we extract $F_0$ out of the Fresnel term and convert the integral into a simple linear function. After a series of derivations, the integral of the specular term can be converted to the following form:

% the integral of the specular term can be converted into a function of $\cos{\theta_o}$ and the surface roughness $\alpha$. $\theta_o$ is the angle between the outcoming direction and the shading normal. 
 
\begin{align}
    \int & \rho_{\mathrm{spec}}(\omega_i,\omega_o) \cos{\theta_i} \mathrm{d} \omega_i  
 %   = & F_0\int \frac{\rho_{\mathrm{spec}}(\omega_i,\omega_o)}{F(\omega_o, \omega_h)}(1-F_\mathrm{c})\cos{\theta_i}\mathrm{d}\omega_i \notag \\ 
 %   + & \int \frac{\rho_{\mathrm{spec}}(\omega_i,\omega_o)}{F(\omega_o, \omega_h)}F_\mathrm{c}\cos{\theta_i}\mathrm{d}\omega_i, \\
    =  F_0 A + B, 
\end{align}

where
\begin{align}
    A & = \int \frac{\rho_{\mathrm{spec}}(\omega_i,\omega_o)}{F(\omega_o, \omega_h)}(1-F_\mathrm{c})\cos{\theta_i}\mathrm{d}\omega_i, \\
    B & = \int \frac{\rho_{\mathrm{spec}}(\omega_i,\omega_o)}{F(\omega_o, \omega_h)}F_\mathrm{c}\cos{\theta_i}\mathrm{d}\omega_i, \\
    F_\mathrm{c} & = (1 - (\omega_o \cdot \omega_h))^5.
\end{align}
The two resulting integrals represent a scale (denoted by $A$) and a bias (denoted by $B$) to $F_0$, respectively. We perform importance sampling on the incident direction vector $\omega_i$, resulting in a 2D lookup table with respect to $\cos{\theta_o}$ and roughness $\alpha$. In our implementation, the resolution of the pre-computation lookup table is 512 $\times$ 512 with 1024 samples per pixel, as shown in Figure~\ref{fig:lut}.

After performing the pre-computation above, we can avoid complex integral of material component $F_\beta$ in real-time inference. We can directly use the albedo map ($c$) and metallic map ($m$) to obtain the integral value of the diffuse term. And use the specular map ($F_0$), NoV map (the dot product of normal and view direction, i.e., $\cos{\theta_o}$) and the roughness map ($\alpha$) to query the pre-computation lookup table to get the integral value of the specular term. The result of adding the two terms is the final material component $F_\beta$.

\begin{figure}
    \centering
    \includegraphics[width=1.0\linewidth]{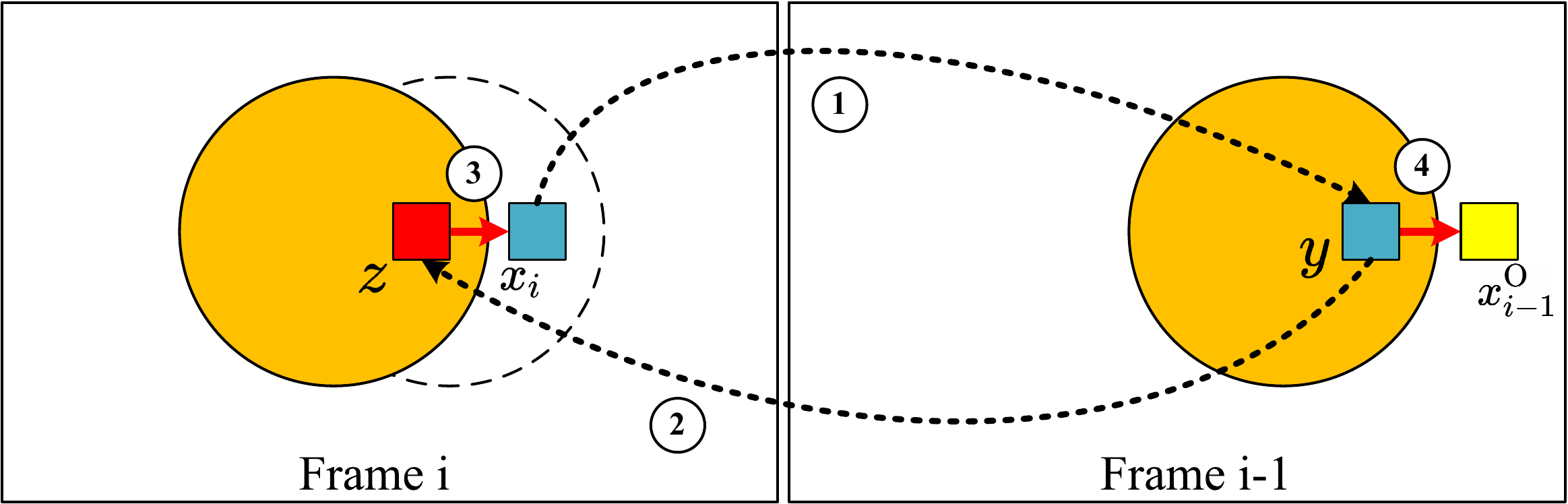}
    \caption{Illustration of the implementation of dual motion vectors for occlusions. For a pixel $x_i$ that is visible now but was occluded in the previous frame at $y$, we find where the occluder $y$ is in the current frame at $z$. Then we find $x_i$'s correspondence $x_{i-1}^\mathrm{O}$ in the previous frame using the relative motion vector from $z$ to $x_i$.}
    \label{fig:dmv}
\end{figure}

\begin{figure}[t]
  \centering
   \includegraphics[width=1.0\linewidth]{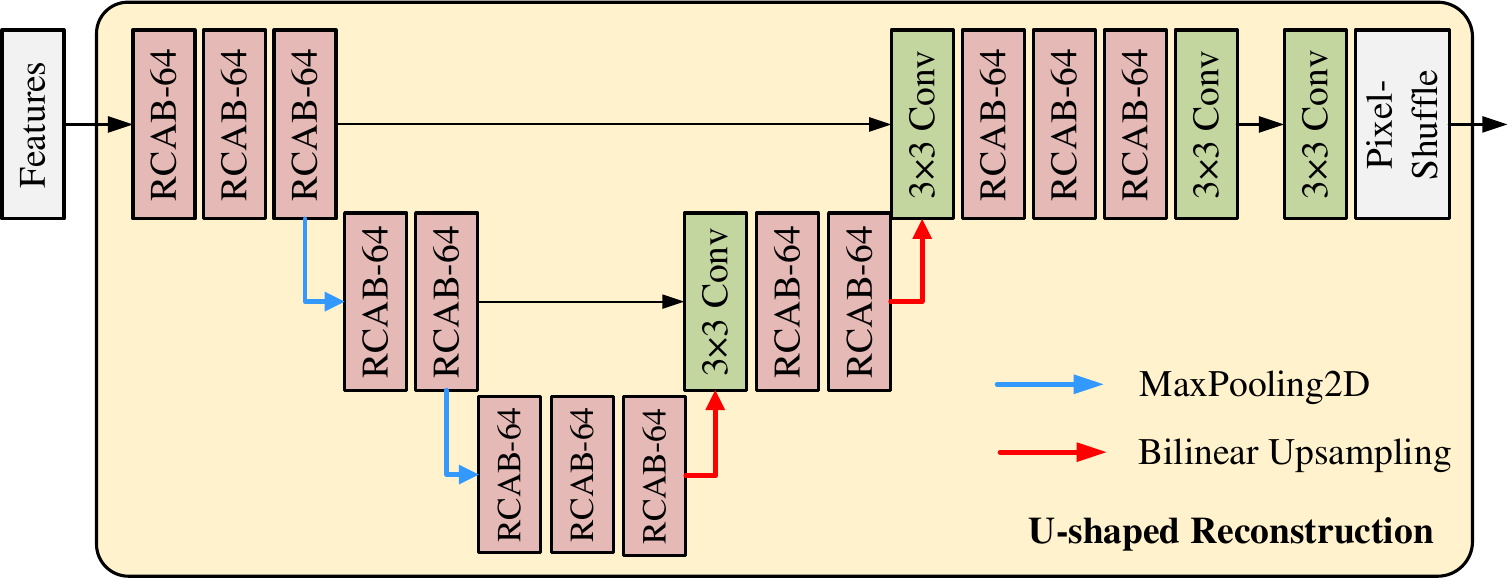}
   \caption{The structure of our U-shaped reconstruction module.}
   \label{fig:unet}
\end{figure}

\section{Details of Dual Motion Vector}
In order to alleviate the ghosting problem caused by warping with the traditional motion vector (TMV) due to object occlusion in dynamic scenes, Zeng et al.~\cite{TRMV} proposed the dual motion vector (DMV), and its implementation process is shown in Figure~\ref{fig:dmv}. First, for the visible point $x_{i}$ in the current frame, find the occluded point $y$ in the previous frame through back projection, where $y-x_i$ is considered as TMV. In the second step, find the point $z$ in the current frame through forward projection from point $y$ in the previous frame. The third step is to get the relative motion vector from point $z$ to point $x_{i}$. Finally, use this relative motion vector to find point $x_{i-1}^\mathrm{O}$ in the previous frame, and set $x_{i-1}^\mathrm{O}-x_i$ as DMV. In our method, we can accurately obtain the occluded area by subtracting DMV with TMV.

\section{Details of Reconstruction Module}
In the frame-recurrent super-resolution module, we use the U-shaped reconstruction module to reconstruct and upsample the intermediate features to obtain the high-resolution lighting components. The structure of the U-shaped reconstruction module is shown in Figure~\ref{fig:unet}. The residual channel attention blocks (RCAB)~\cite{RCAN} in the module have been widely used in super-resolution due to the ability to improve reconstruction quality, and we choose the U-shaped structure to connect them in order to reduce the network computation as much as possile. We use max-pooling layer for downsampling, bilinear interpolation for upsampling and channel-wise connections for preserving the shallow features. Finally, the pixel-shuffle operation~\cite{Shi:2016:ESPCN} is used for upsampling reconstruction. % All convolutional layer output channels in the reconstruction module are 64, although the other two modules (radiance demodulation and reliable warping module) are 32.

\section{Experiments}

\subsection{Datasets}
\label{sec:sup_datasets}
\begin{figure}
    \centering
    \includegraphics[width=1.0\linewidth]{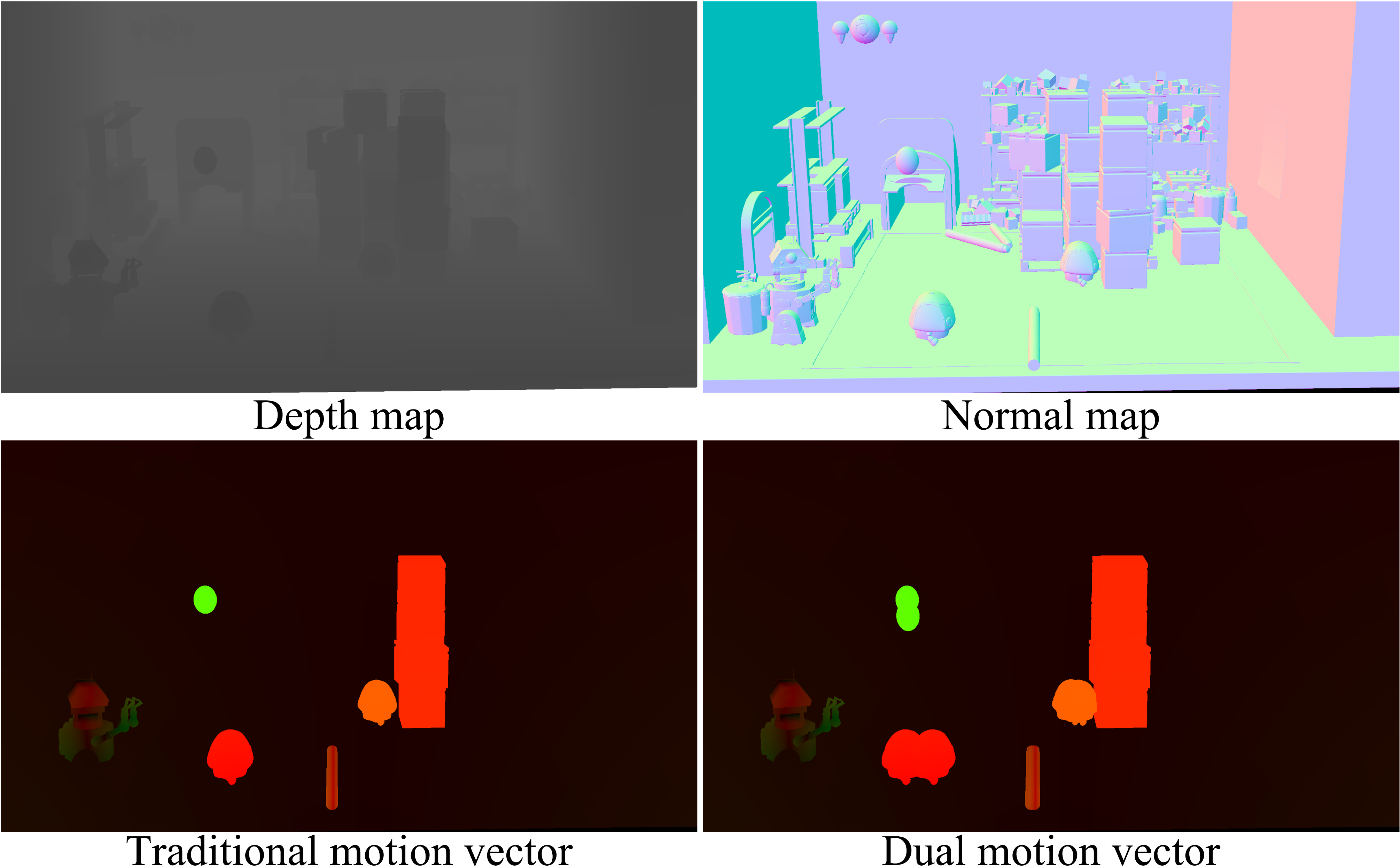}
    \caption{Additional network inputs of the Pica scene.}
    \label{fig:gbuffer}
\end{figure}

\begin{figure*}
    \centering
    \includegraphics[width=1.0\textwidth]{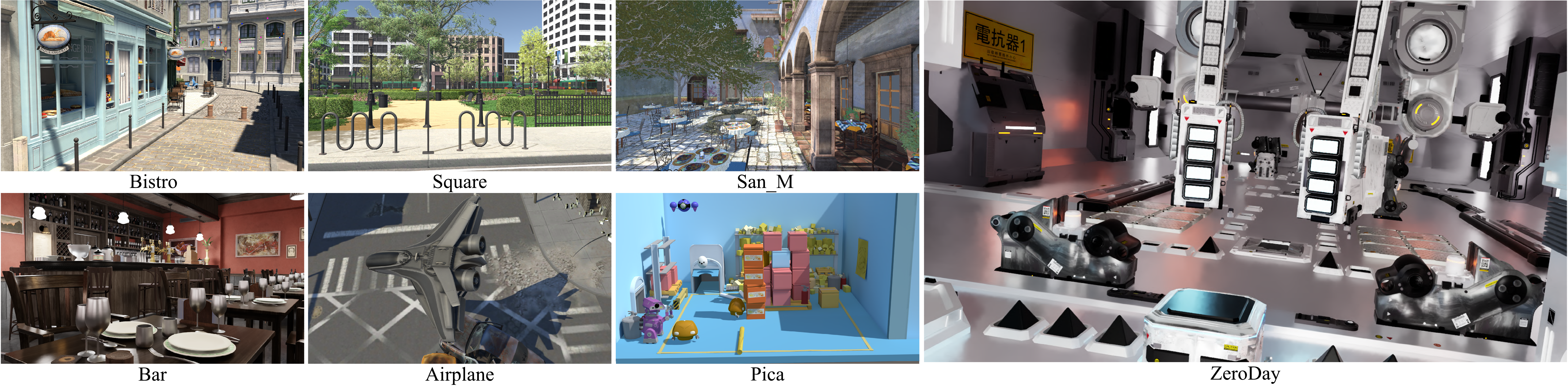}
    \caption{Example images of seven scenes.}
    \label{fig:dataset}
    \vspace{-0.8em}
\end{figure*}

We use the Unity~\cite{unity} rendering engine to generate our dataset. We select seven representative scenes, namely Bistro~\cite{Bistro}, Square~\cite{ORCANVIDIAEmeraldSquare}, San Miguel (San\_M)~\cite{McGuire:2017:Data}, Bar~\cite{Bistro}, ZeroDay~\cite{ZeroDay}, Airplane and Pica. The example images are shown in Figure~\ref{fig:dataset}. Similar to previous work~\cite{NSRR}, we uniformly distribute different fast-moving cameras in each scene to generate multiple sequences of 100 frames each, containing as different objects and materials as possible to enhance diversity. We randomly divided the training, validation and testing datasets from these sequences, and the exact number of sequences is shown in Table~\ref{tab:data}. We also generate a set of rendering G-buffers as the additional inputs to the network. An example is shown in Figure~\ref{fig:gbuffer}. 

\subsection{Implementation Details}
\label{sec:sup_impl}
We demodulate the LR radiance into the lighting component with the LR and HR material G-buffer generated in the deferred rendering pipeline, as described in Section~\ref{sec:sup_demod}. Then, the light component, together with the LR depth, normal, motion vector, and dual motion vector into the network, is fed into the network.

In our super-resolution neural network, we set the convolutional output channel number as 32 in the radiance demodulation and reliable warping modules. The convolutional output channel number in the frame-recurrent reconstruction module is set as 64 (except for the last channel number is 3 $\times$ \textit{s} $\times$ \textit{s} for the pixel-shuffle operation, where \textit{s} is the SR factor). The breakdown of our network is shown in Table~\ref{tab:sup_para}.

The output of the network -- the reconstructed HR light component, is modulated with the HR material component to form the current output frame and aids the reconstruction for the next frame.

% The diffuse map and specular map are the interpolation results of albedo and Fresnel coefficient ($F_0$) using metallic map, respectively.

\begin{table}
  \centering
  \small
  \caption{The sequence number of training, validation and testing dataset for each scene. Each sequence contains 100 frames.}
  \label{tab:data}
  \begin{tabular}{lcccc}
    \toprule
     Scene & \tabincell{c}{Training\\Sequences} & \tabincell{c}{Validation\\ Sequences} & \tabincell{c}{Testing \\Sequences}\\
    \midrule
     Bistro     &   54  &   6  &  12 \\
     San\_M      &   54  &   6  &  6  \\
     Square     &   42  &   6  &  6  \\
     Bar        &   45  &   3  &  6  \\
     ZeroDay    &   24  &   3  &  3  \\  
     Pica       &   30  &   3  &  3  \\
     Airplane   &   24  &   3  &  3  \\
  \bottomrule
\end{tabular}
\end{table}

\begin{table}
  \centering
  \small
  \caption{The parameters and GFLOPs of each module.}
  \label{tab:sup_para}
  \begin{tabular}{l|c|c|c}
    \hline
    \multicolumn{2}{c|}{}  & Params (K)  & GFLOPs   \\
    \hline
    \multicolumn{2}{c|}{Radiance Demodulation}  & 2.05 & 0.26 \\
    \hline
    \multicolumn{2}{c|}{Reliable Warping}       & 9.34 & 1.20 \\
    \hline
    \multirow{3}*{\tabincell{c}{Frame-Recurrent\\Reconstruction}} & First conv     & 13.86  & 1.79 \\
    \cline{2-4}     
                                                                  & ConvLSTM & 442.62 & 57.33 \\
    \cline{2-4}
                                                                  & Ushaped-Net     & 1142.92 & 84.78 \\
    \hline
    \multicolumn{2}{c|}{Total} & 1610.79 & 145.36 \\
  \hline
\end{tabular}
\end{table}

\begin{table}
  \small
  \centering
  \caption{Comparison between our method and EDSR with four error measurements on the Bistro scene. The SR factor is set as 4$\times$4.}
  \label{tab:edsr}
  \begin{tabular}{l|c|c|c|c}
    \hline
     & PSNR(dB) & SSIM & LPIPS $\downarrow$ & VMAF  \\
     \hline
     EDSR & 24.08 & 0.7625  & 0.327 & 33.25   \\
     Ours   & \textbf{26.43} & \textbf{0.8739}  & \textbf{0.141} & \textbf{53.82}   \\
  \hline
\end{tabular}
\end{table}

\begin{table}
  \centering
  \small
  \caption{Reconstruction quality versus the SR factor on the Bistro scene with the target resolution set as 1920 $\times$ 1080. The metrics are averaged over all test data (1200 frames).}
  \label{tab:factor}
  \begin{tabular}{lccc}
    \toprule
     SR factor & 2 $\times$ 2 & 4 $\times$ 4 & 6 $\times$ 6 \\
    \midrule
     PSNR &   30.11  &  26.43  &  25.18    \\
     SSIM &   0.9405 &  0.8739 &  0.8349    \\
  \bottomrule
\end{tabular}
\end{table}

\subsection{Comparison of Qualitative Results}
\begin{figure*}
    \centering
    \includegraphics[width=1.0\textwidth]{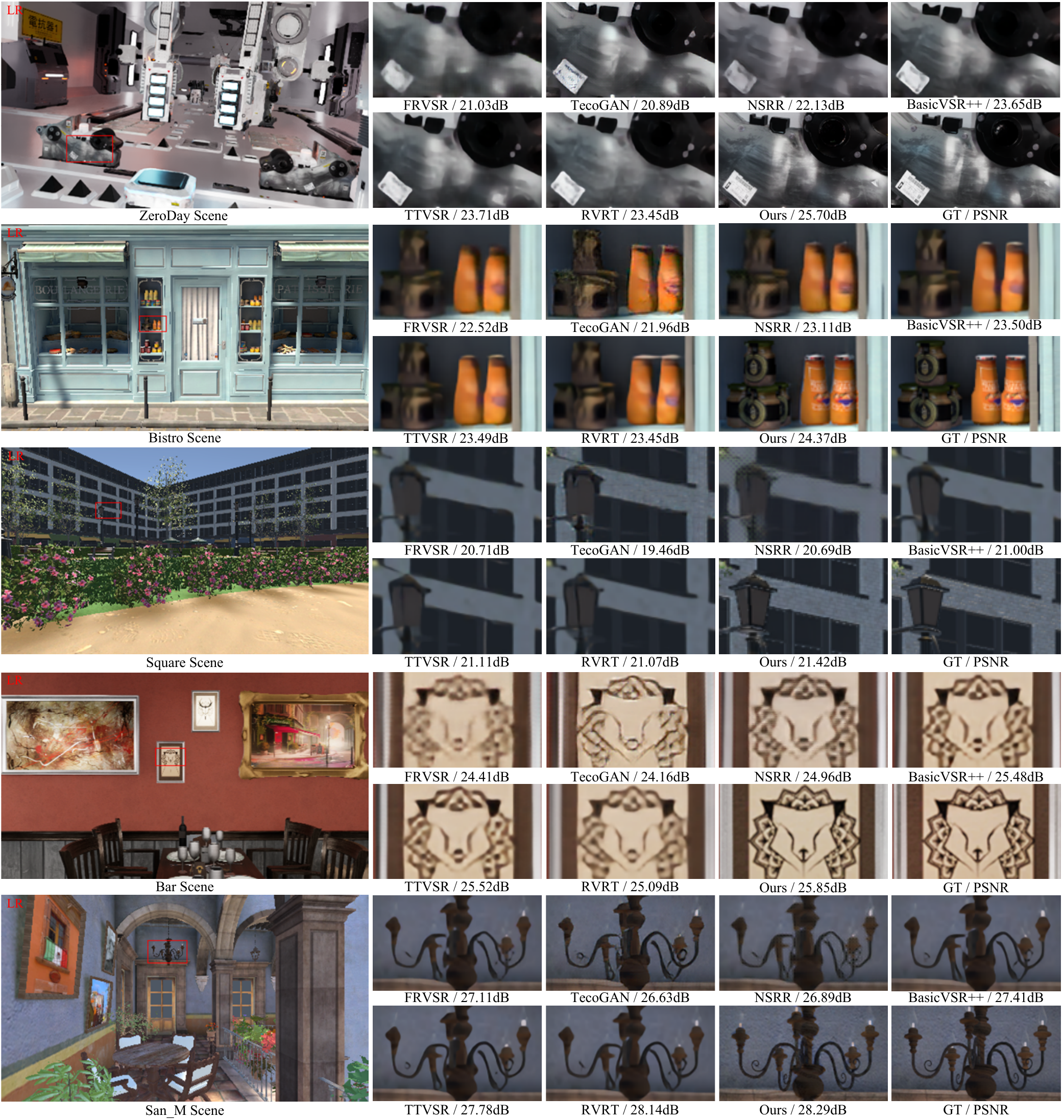}
    \caption{ Comparison among our method, FRVSR~\cite{FRVSR}, TecoGAN~\cite{tecoGAN}, NSRR~\cite{NSRR}, BasicVSR++~\cite{chan2022basicvsr++}, TTVSR~\cite{liu2022ttvsr} and RVRT~\cite{liang2022rvrt}. The target resolution is set as 1920 $\times$ 1080 and the upsampling ratio is set as 4 $\times$ 4.}
    \label{fig:quality1}
    \vspace{-0.8em}
\end{figure*}

\begin{figure*}
    \centering
    \includegraphics[width=1.0\textwidth]{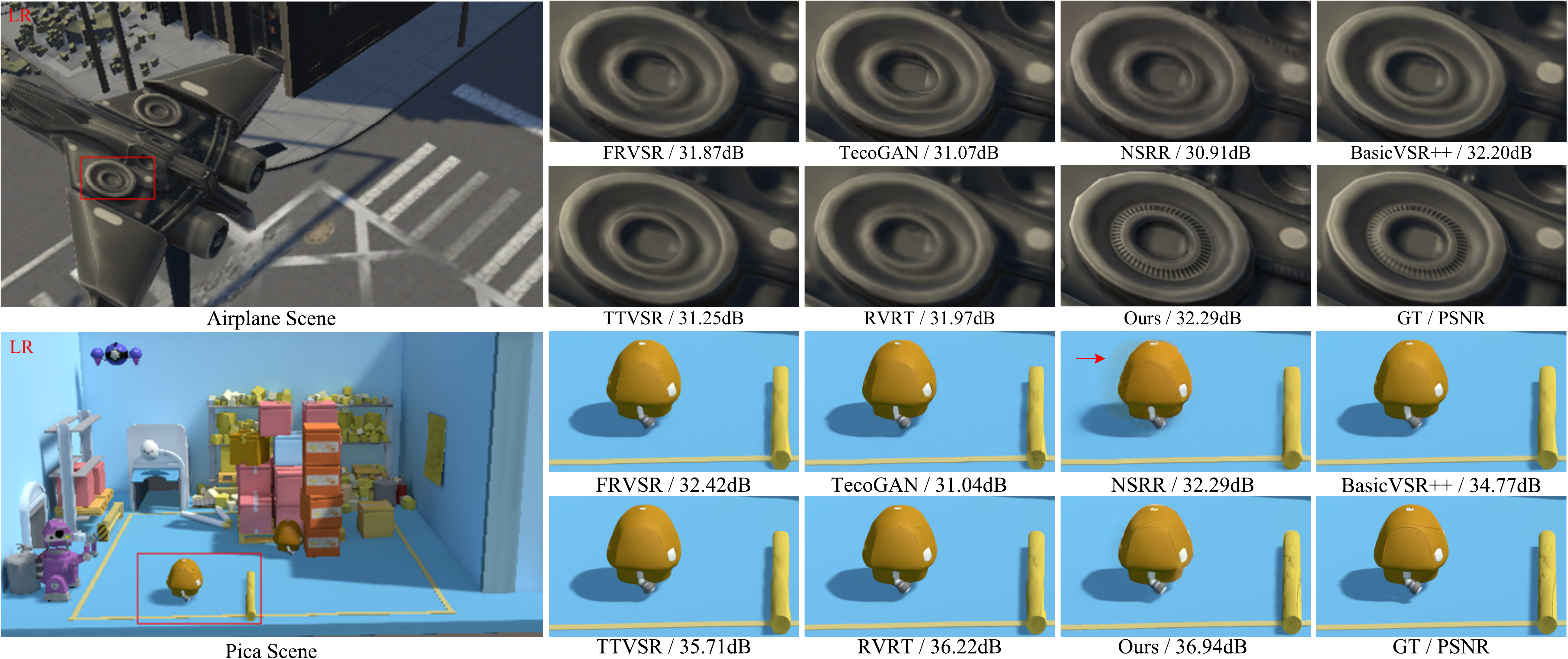}
    \caption{ Comparison among our method, FRVSR~\cite{FRVSR}, TecoGAN~\cite{tecoGAN}, NSRR~\cite{NSRR}, BasicVSR++~\cite{chan2022basicvsr++}, TTVSR~\cite{liu2022ttvsr} and RVRT~\cite{liang2022rvrt}. The target resolution is set as 1920 $\times$ 1080 and the upsampling ratio is set as 4 $\times$ 4.}
    \label{fig:quality2}
\end{figure*}

We provide additional qualitative comparisons of the results on seven scenes, as shown in Figure~\ref{fig:quality1} and Figure~\ref{fig:quality2}. From the results, our method not only produces results with richer texture details (Bistro Scene) but also recovers the view-dependent highlights (ZeroDay scene) well. It can also be seen from the Pica scene that our method successfully eliminates the obvious ghosting problem in NSRR~\cite{NSRR}. Furthermore, as can be seen from the video in the supplementary material, our method has better temporal stability compared to other methods.

\subsection{Comparison of Generalization Ability}
\begin{figure*}
    \centering
    \includegraphics[width=1.0\textwidth]{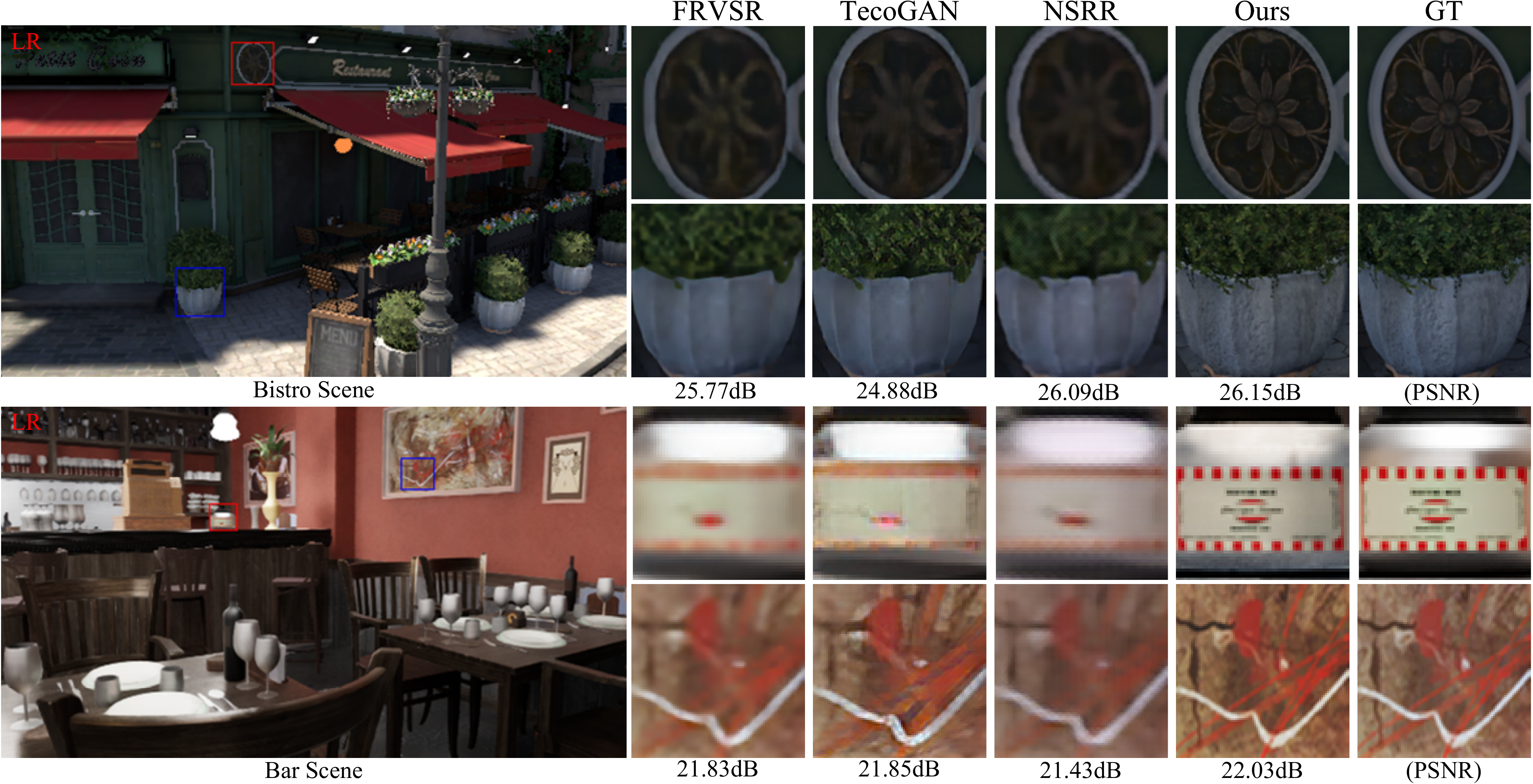}
    \caption{Comparison of generalization ability among our method, FRVSR~\cite{FRVSR}, TecoGAN~\cite{tecoGAN} and NSRR~\cite{NSRR}. The target resolution is set as 1920 $\times$ 1080 and the upsampling ratio is set as 4 $\times$ 4.}
    \label{fig:general}
\end{figure*}

We have compared the generalization results of FRVSR~\cite{FRVSR}, TecoGAN~\cite{tecoGAN} and NSRR~\cite{NSRR} quantitatively in the main paper, and we also provide a comparison of qualitative results in Figure~\ref{fig:general}. It can be seen from the results that other methods cause excessive blurring of details. Although TecoGAN can get a slightly sharp result, it is still quite different from GT. However, our method can still preserve complex texture details, which proves that our method has generalization ability.

\subsection{Varying SR Factors Results}
\begin{figure*}
    \centering
    \includegraphics[width=1.0\textwidth]{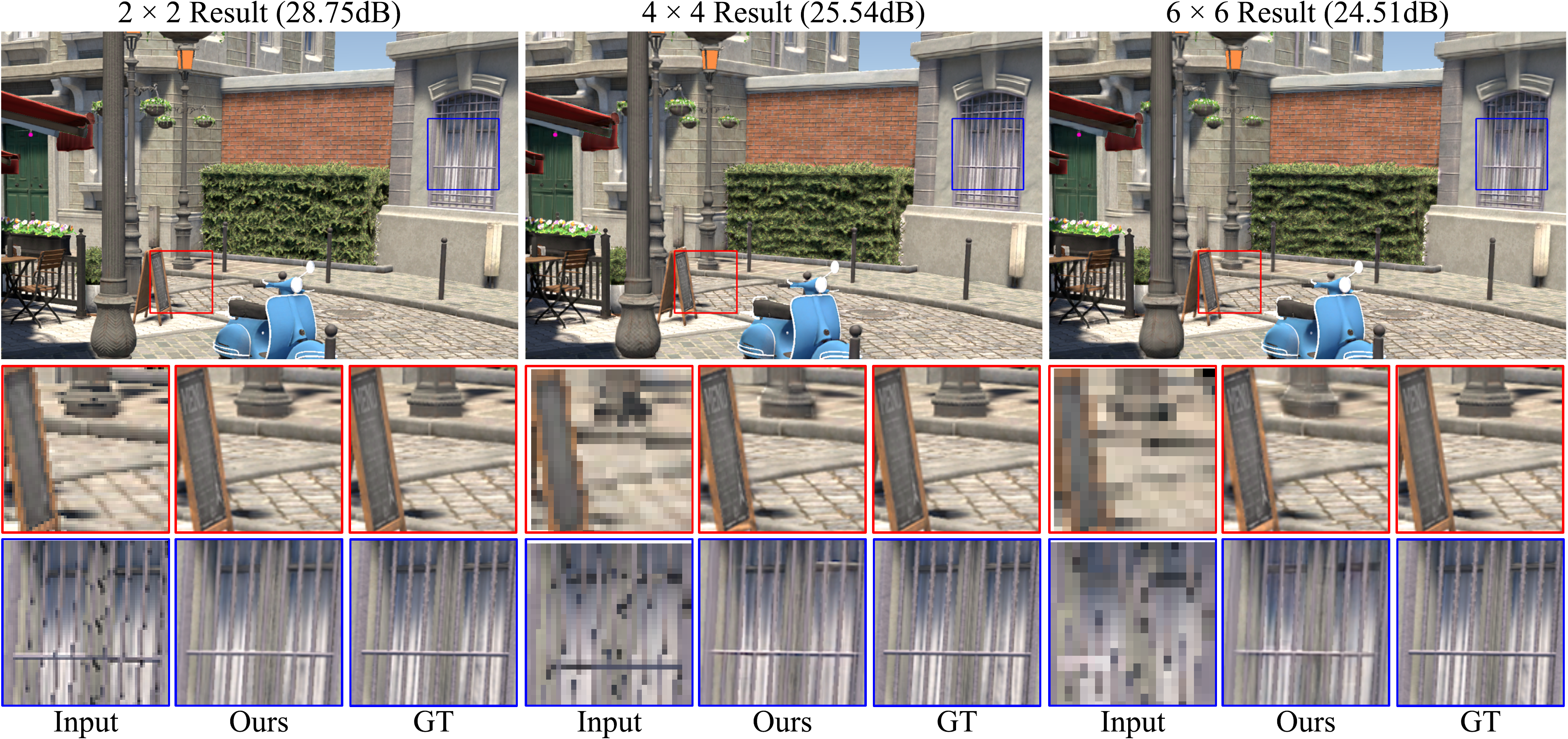}
    \caption{Reconstruction quality versus the SR factor on the Bistro scene with the target resolution set as 1920 $\times$ 1080. The results of PSNR are tested on the single full-resolution image.}
    \label{fig:SRF_comp}
\end{figure*}

Table~\ref{tab:factor} and Figure~\ref{fig:SRF_comp} show the quantitative and qualitative reconstruction results under different SR factors, respectively. We keep the target resolution (1920 $\times$ 1080) the same and modify the input image resolution according to the SR factors. As the SR factor increases, the error becomes larger, since the SR reconstruction becomes more difficult. Our method can still preserve rich texture details thanks to the radiance demodulation module.

\subsection{Comparison with SISR Method}
\begin{figure}
    \centering
    \includegraphics[width=1.0\linewidth]{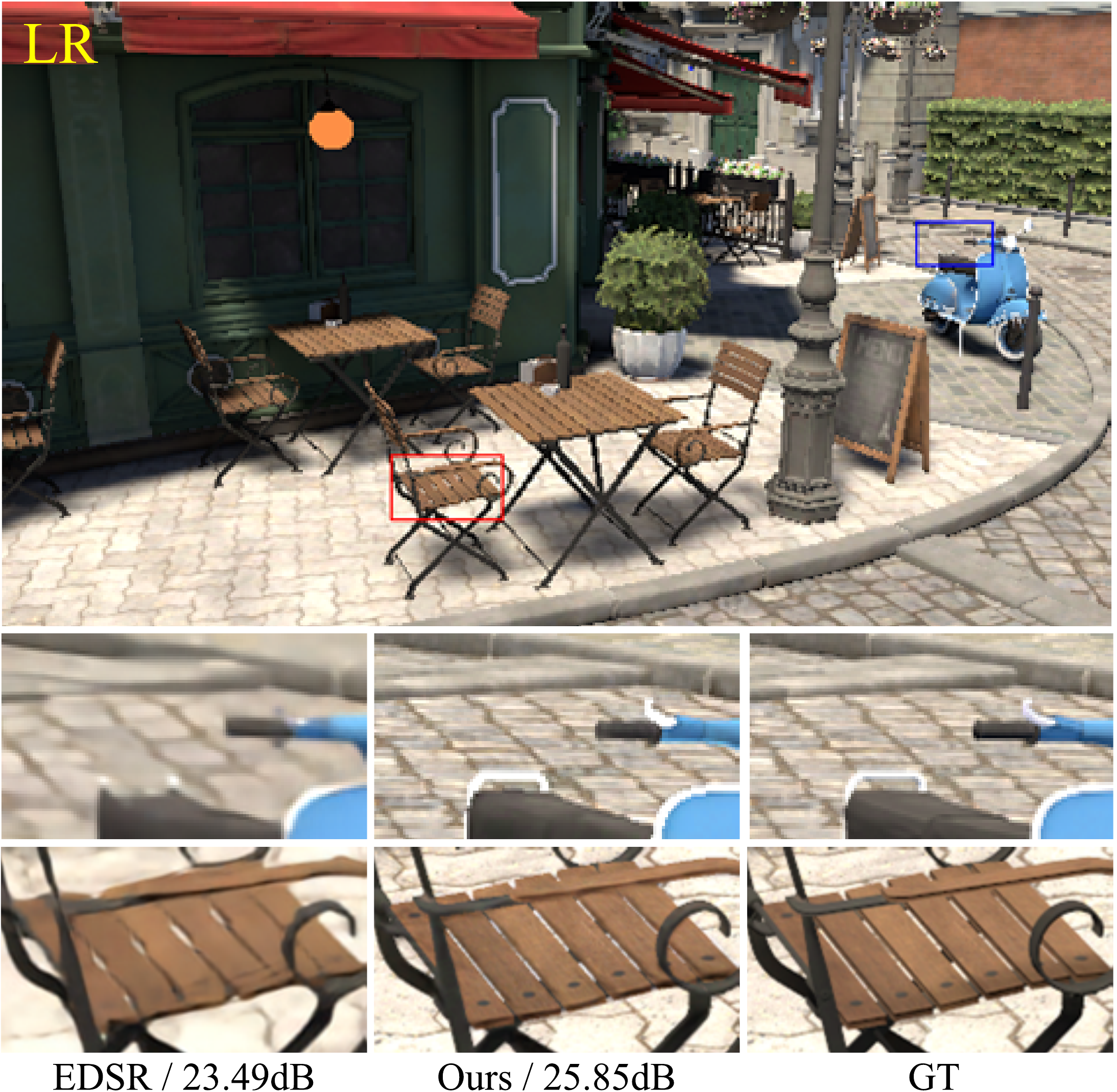}
    \caption{Comparison with EDSR on the Bistro scene. The target resolution is set as 1920 $\times$ 1080 and the SR factor is set as 4 $\times$ 4.}
    \label{fig:edsr}
\end{figure}

We compare our method with a SISR method, i.e., EDSR~\cite{EDSR}, in Table~\ref{tab:edsr} and Figure~\ref{fig:edsr}. Our method shows higher quality than EDSR on all metrics, and the details are better preserved. Furthermore, EDSR leads to poor temporal stability, as demonstrated in the video, since they do not consider the consistency between consecutive frames.

\subsection{Limitations} 
\begin{figure}
    \centering
    \includegraphics[width=1.0\linewidth]{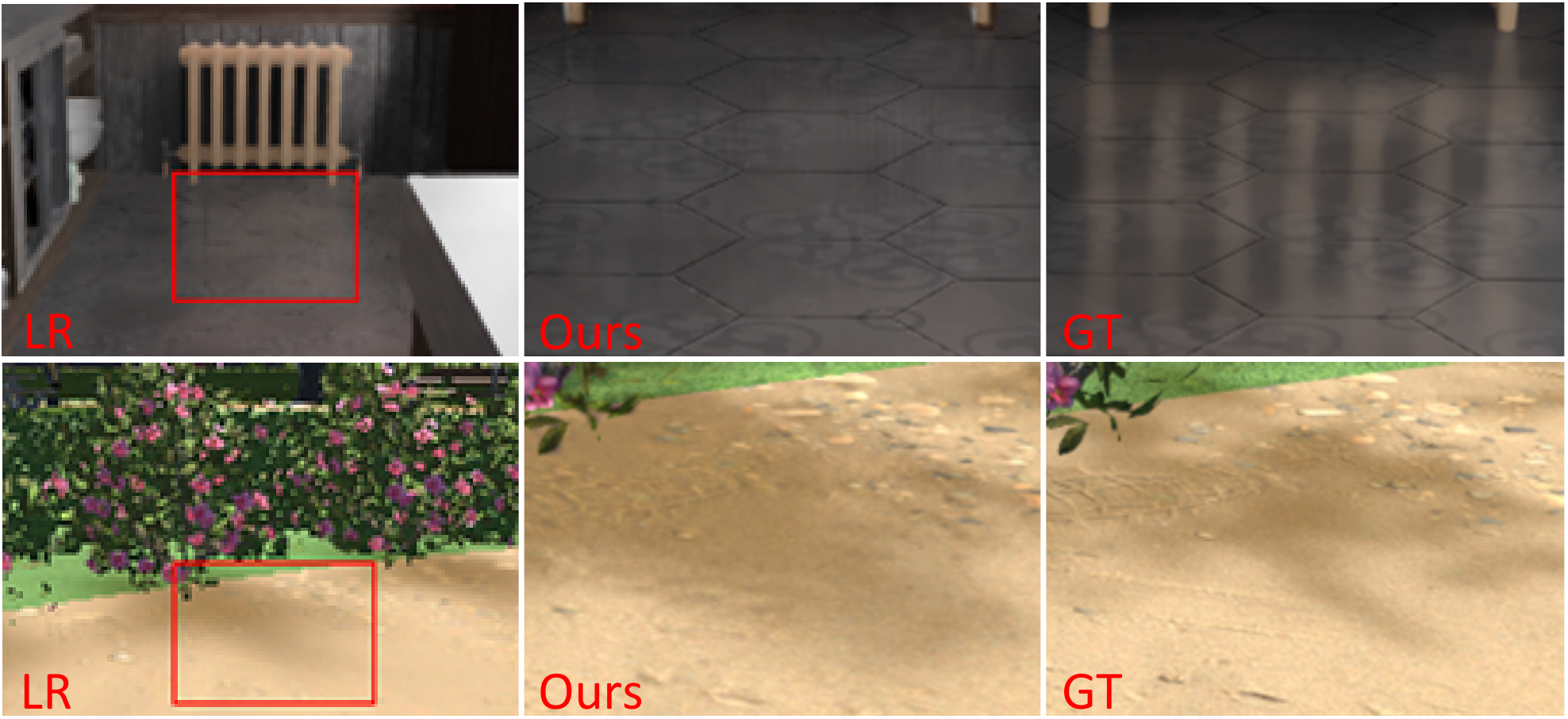}
    \caption{Failure cases on high-frequency indirect reflections and shadow boundaries. }%Our method still has difficulty recovering high-frequency indirect reflections and shadow boundaries finely when the light moves fast.}
    \label{fig:limit}
\end{figure}

Although our method produces high-fidelity results in most scenarios, we have still identified some limitations, including complex indirect reflections and moving shadows, which are known to be challenging, as shown in Figure~\ref{fig:limit}. These high-frequency effects are due to the lighting rather than the material component. Therefore, our method shows subtle benefits.

% \subsection{Fair Comparison with BasicVSR++}
% \revise{We modify BasicVSR++~\cite{chan2022basicvsr++} for a fair comparison. Specifically, we first apply the same G-buffer information and radiance demodulation as our method to BasicVSR++. And we only keep its forward feature propagation module (denoted as BasicVSR++ (Rendering)), because we can only obtain historical information during real-time rendering, and BasicVSR++ is a bidirectional method (using both future and historical information). We also train a lighter version (denoted as BasicVSR++ (Rendering\_S)) by reducing the number of channels by half. The comparison results are shown in Table~\ref{tab:fair}. Our method achieves comparable quality with 1/3 the number of parameters and 1/6 the running time of BasicVSR++ (Rendering), and surpasses BasicVSR++ (Rendering\_S) in both quality and performance.}

%------------------------------------------------------------------------

{\small
\bibliographystyle{ieee_fullname}
\bibliography{supplement}
}